\documentclass[11pt]{article}

\usepackage{amssymb, amsthm}
\usepackage{latexsym}
\usepackage{amsmath}
\usepackage[ansinew]{inputenc}
\usepackage{hyperref}

\numberwithin{equation}{section}

\newtheorem{theo}{Theorem}[section]
\newtheorem{prop}[theo]{Proposition}

\newtheorem{lem}[theo]{Lemma}
\theoremstyle{definition}
\newtheorem{defi}[theo]{Definition}
\newtheorem{bsp}[theo]{Example}

\begin{document}
\begin{titlepage}

\date{\today}
\begin{center}
{\bf\Large Geometric and deformation quantization}

\vskip1.5cm

{\normalsize Christoph Nölle}

\vskip1cm

{\it Institut für Theoretische Physik, Leibniz Universität Hannover, \\
 Appelstraße 2, 30167 Hannover, Germany}\\
 email: noelle@math.uni-hannover.de

\end{center}

\vskip1.5cm

\begin{abstract}
 We present a simple geometric construction linking geometric to
 deformation quantization. Both theories depend on some apparently
 arbitrary parameters, most importantly a polarization and a symplectic connection,
 and for real polarizations we find a compatibility condition
 restricting the set of admissible connections. In the special case when phase space is a cotangent bundle
 this compatibility condition has many solutions, and the
 resulting quantum theory not only reproduces the well-known
 geometric quantization scheme, but also allows to quantize all
 interesting observables. For Kähler manifolds there is no
 compatibility condition, but a canonical choice for the parameters. The explicit form of the observables
 however remains undetermined.

\end{abstract}
\end{titlepage}
\setcounter{tocdepth}{1}

 \tableofcontents

\section{Introduction}
 It is often stated that the problem of how to quantize a symplectic or Poisson
 manifold $M$ (phase space in physical terms) has been solved by
 deformation quantization, in particular by the work of Fedosov
 (for symplectic manifolds) \cite{fedosov:deformationart,fedosov:deformationbook}
  and Kontsevich (for the more general Poisson
 manifolds) \cite{kontsevich_deformPoisson}. These constructions give the most general method to deform
 the pointwise product on $C^\infty(M)$ into a noncommutative
 product $\ast$, such that certain physical requirements are
 satisfied. The product $\ast$ corresponds to composition of operators in the ordinary
 framework of quantum theory on a flat phase space, and is called
 (Groenewold-)Moyal product there
 \cite{Groenewold_QM,moyal:original}.

 What is missing in this formalism is the construction of a Hilbert
 space on which the resulting algebra acts, i.e. the space of
 states. This is not really a problem from a conceptual point of view, because one can
 define states in a purely algebraic framework as positive linear
 functionals on the algebra. In the present situation this is not completely straightforward,
 as deformation quantization usually does not produce
 well-behaved algebras, like $C^\ast$ algebras, but it might be possible to circumvent
 this problem. Once one has the states defined as functionals,
 one can also define representations of the algebra on Hilbert
 spaces, using the GNS construction \cite{Waldmann_StatesandReps}.

  However, classical quantum mechanics on flat space features one
 more physical property, which cannot easily be incorporated into
 the deformation framework. The wave functions there have a
 probability interpretation, e.g. in the position space
 representation $|\psi(q)|^2$ is a density in space, whose
 integral over some region $\Omega \subset \mathbb R^n$ gives the
 probability to find the particle in $\Omega$. Another example is
 the Fock space representation, where $|\psi_n|^2$ gives the
 probability that $n$ quanta or particles will be found in the state
 $\psi$. In principle this information can be extracted from the
 expectation values of suitable observables, e.g. $\int_\Omega
 |\psi(q)|^2 dq=\langle\hat \chi_\Omega\rangle_\psi$, where
 $\chi_\Omega$ is the characteristic function of $\Omega\times\mathbb R^n$ ($\mathbb R^n=$ momentum space),
 and $\hat \chi_\Omega$ its
 associated operator. The advantage of the classical formalism is
 really that the calculation of these expectation values becomes
 simple for a large class of observables.

 For a long time the Hilbert space had therefore been considered as
 an essential ingredient of the theory, and the attempts to define
 quantization rigorously on curved phase spaces focused on its
 construction. These ideas culminated in the theory of geometric quantization,
 which provides a method for the construction of a Hilbert space consisting of
 wave functions with the same probability interpretation as in the
 flat case \cite{woodhouse:quantisierung, Ali_QuantMeths}.
 This works for symplectic phase spaces, to which we restrict our attention in this article.
 It is however not possible to give a general
 construction of all the observables in this framework, despite some
 promising ideas.

  It would therefore be desirable to unify these two constructions such that
 one obtains a representation of the deformed algebra $\mathcal A$ on the Hilbert
 space $\mathcal H_P$ of geometric quantization. We will show in the paper that the
 geometrical structures appearing in the two theories are indeed
 closely related, that $\mathcal A$ has a canonical
 representation on a Hilbert space different from but closely related to $\mathcal H_P$,
 and that this representation can be carried over to $\mathcal H_P$
 in an obvious way. However, one assumption
 has to be made, which is related to the parameters occurring in the
 separate theories.

  This is not surprising; neither geometric nor deformation quantization
 are uniquely determined by the symplectic manifold $(M,\omega)$, but
 depend on further structures. In geometric quantization these are
 essentially a polarization of the tangent bundle, a metaplectic
 structure, and a so called prequantum line bundle.
 Deformation quantization enforces the choice of a symplectic
 connection on $M$, although it can be shown that different
 connections lead to isomorphic algebras. We will see that
 deformation quantization is also closely related to metaplectic
 structures, which gives the first hint at some relation between the two
 theories. Our assumption essentially gives a compatibility condition between
 the polarization and symplectic
 connection.

  We will study in detail the case of cotangent bundles $T^\ast Q$,
 equipped with their canonical symplectic form, and show that the
 compatibility condition has many solutions in this case, even after
 fixing the polarization. Note that (at least at first sight) it would be much more desirable
 to find a unique solution, in order to decrease the number of
 apparently arbitrary parameters occurring in the quantization
 process. For manifolds with a totally complex polarization the
 compatibility condition does not even impose any restrictions on
 the connection.

 There are two more advantages of geometric over deformation
 quantization. The latter often produces an infinite series for the
 observables, whose convergence properties are poorly understood,
 whereas the geometric operators are quite simple. What we
 will find in the case of cotangent bundles is that although
 our final quantum operators are constructed from the formal deformation
 operators, they are actually perfectly well defined for a large class of
 functions.

 Further, there may exist inequivalent irreducible representations
 of the quantum algebra, and the different physical content of these
 representations often becomes at least partly obvious in the geometric
 quantization scheme, whereas it may be hard to extract this
 information from a state defined as a functional on the algebra.

 \bigskip

  Our notation will be a bit sloppy at times. Instead of dealing with
 triples $\mathcal S(\mathbb R^n)\subset L^2(\mathbb
 R^n)\subset \mathcal S'(\mathbb R^n)$, where $\mathcal S$ denotes
 Schwartz functions, and $\mathcal S'$ the dual space consisting of
 tempered distributions, we only work with the Hilbert space $L^2(\mathbb R^n)$, and
 pretend e.g. that it contains a delta-function. This is in order to keep the notation simple,
 but it should not be difficult to translate the results into rigorous statements.

  In the first chapter we present a construction of
 the metaplectic group as a subgroup of the Weyl algebra. The
 conventional approach is to work in a fixed representation of the
 latter, such that the metaplectic group action becomes unitary.
 This is not really a restriction, as all irreducible
 representations are equivalent, by the Stone-von Neumann theorem.
 In order to demonstrate the importance of polarizations in the
 representation theory of the Weyl algebra and metaplectic group
 (thus giving a purely mathematical argument why geometric and deformation
 quantization belong together) we
 choose to work in a somewhat more formal setting here, and
 construct the representations only at the end of the section.
\parindent=0cm

\section{The Metaplectic Group}\label{Section:MetaRep}
\subsection{Formal construction}
 Let $(V,\omega)$ be a real, $2n$ dimensional, symplectic vector
 space, and
  $$\text{Sp}(V,\omega)=\{S \in \text{ Aut}(V)\ |\
 \omega(Sv,Sw)=\omega(v,w)\ \forall v,w\in V\}$$
 the symplectic group of $V$, with Lie algebra
  \begin{equation}\label{sympLA}
        \mathfrak{sp}(V,\omega)=\{A\in \text{End}(V)\ |\
  \omega(Av,w)+\omega(v,Aw)=0\ \forall v,w\in V\}.
  \end{equation}
 Consider the symmetric tensor algebra Sym($V^\ast_\mathbb C)=
 \bigoplus_{k\geq 0} $Sym$_k (V^\ast_\mathbb C)$ on the dual $V^\ast$. In the
 following we will simply write $vw=wv$ for the symmetrized tensor product $v\otimes_Sw$, if $v,w\in
 $ Sym$(V^\ast_\mathbb C)$. Symmetrization is understood with factors, e.g.
 $vw=\frac 12(v\otimes w+w\otimes v)$ for $v,w\in V^\ast$.
 Now we introduce a new product $\circ$ on Sym($V^\ast_\mathbb C)$ by
 dividing out from the full tensor algebra $\mathcal
 T(V^\ast_\mathbb C)$ the ideal generated by elements $v\otimes w-w\otimes v
 - i\hbar \omega^{-1}(v,w)$, where $v,w\in V^\ast_\mathbb C$.
 For the time being $\hbar$ can be considered as a positive real number.
 The resulting space is indeed isomorphic as a vector space to
 Sym$(V^\ast _\mathbb C)$, as any of its elements can
 be brought to a symmetric form by reordering. In particular we get
  \begin{equation}\label{WeylProd}
   v\circ w= vw + \frac {i\hbar}2\omega^{-1}(v,w) ,\quad \forall
   v,w\in V^\ast_\mathbb C,
  \end{equation}
 as well as the 'canonical commutation relation'
  \begin{equation}\label{CanComRel}
        [v,w]_\circ= i\hbar \omega^{-1}(v,w),\quad \forall v,w\in
        V^\ast_\mathbb C.
  \end{equation}

\begin{defi}
 The Weyl algebra associated to $(V,\omega)$ is
 $W(V^\ast) :=$ Sym($V^\ast_\mathbb C,\circ)$.
\end{defi}

 Choose a basis $\{e_1,\ldots,e_{2n}\}$ of $V$, and let $\{y^1,\ldots,y^{2n}\}$
 be the dual basis of $V^\ast\subset W(V^\ast)$. The Weyl operators
 $W(a)\in W(V^\ast)$ are defined as
  \begin{equation}\label{WeylOpsDefi}
        W(a) = \exp_\circ\Big\{\frac i\hbar \omega_{ij}a^i  y^j\Big\},
        \qquad a=a^ie_i \in V,
  \end{equation}
 and a formal application of the Baker-Campbell-Hausdorff formula
 shows that
 \begin{align}
    W(a)\circ W(b)= W(a+b)\exp\Big[\frac i{2\hbar}\omega(a,b)\Big].
 \end{align}
 Therefore the Weyl operators $W(a)$ form a group $H(V,\omega)$, called Heisenberg
 group, which can be defined abstractly as $V\times \mathbb R$ with composition
  \begin{equation}
    (v,s)(w,t) = \big(v+w,s+t+\textstyle {\frac 1{2}}\omega(v,w)\big).
  \end{equation}
 It is a central extension of the translation group $V$. We define a map
 \begin{equation}\label{SpMpAlgIsom}
   dU:\mathfrak{sp}(V,\omega) \rightarrow
   \text{Sym}_2(V^\ast_\mathbb C),\ A \mapsto -\frac
   i{2\hbar}\omega_{ij}{A^j}_k y^iy^k,
 \end{equation}
 and observe that for $A\in\mathfrak{sp}(V,\omega)$ the expression
 $\omega_{ij}{A^j}_k$ is symmetric in $i$ and $k$, so that
 (\ref{SpMpAlgIsom}) can also be written as $dU(A)= -\frac
   i{2\hbar}\omega_{ij}{A^j}_k y^i\circ y^k$, which simplifies the
   proof (that we skip) of
\begin{lem}
 $dU:\mathfrak{sp}(V,\omega) \rightarrow W(V^\ast)$ is a Lie algebra homomorphism,
 i.e. for any $A,B\in \mathfrak{sp}(V,\omega)$ the relation
  \begin{equation}
   [dU(A),dU(B)]_\circ = dU([A,B])
  \end{equation}
 holds.
\end{lem}
\begin{defi}
 The metaplectic Lie algebra is $\mathfrak{mp}(V,\omega):=
 dU\big(\mathfrak{sp}(V,\omega)\big)$, and the metaplectic group
 Mp$(V,\omega) =\exp_\circ \big(\mathfrak {mp}(V,\omega)\big)$.
\end{defi}
 As $dU$ is injective, $\mathfrak{mp}(V,\omega)$ is isomorphic to
 $\mathfrak{sp}(V,\omega)$, but the associated
 Lie groups are not. It is known that the metaplectic group forms a two-fold covering group of
 Sp$(V,\omega)$, but it is not the universal cover, as
 $\pi_1($Sp$(V,\omega))=\mathbb Z$
 \cite{folland:harmonic_analysis_in_ps}. Any faithful
 representation of Mp$(V,\omega)$ must be infinite-dimensional,
 although it is a finite-dimensional Lie group. To $S=e^A\in $ Sp$(V,\omega)$ we associate
 $U(S)\in $ Mp$(V,\omega)$ through $U(S)=e^{dU(A)}$, and note that $U(S)$ is
 only defined up to a sign by $S$.

\begin{lem}\label{MetaProperties}
 For $A\in \mathfrak{sp}(V,\omega)$ and $S\in Sp(V,\omega)$ the
 following relations hold
 \begin{enumerate}
   \item $[dU(A),y^i]_\circ = -{A^i}_jy^j$
   \item $U(S)\circ y^i\circ U(S)^{-1} = {(S^{-1})^i}_jy^j$
   \item \label{MetaGroupDefi}$W(Sa) = U(S)\circ W(a)\circ U(S)^{-1},\qquad \forall a\in V$
 \end{enumerate}
\end{lem}
\begin{proof}\
 \begin{enumerate}
   \item is a direct consequence of the definition of $dU$ and the
 commutation relation (\ref{CanComRel}).
   \item follows from 1, and
   \item from 2:
    \begin{align*}
      W(Sa)&= \exp_\circ \Big\{\frac i\hbar \omega_{ij}{(Sa)^i}
         y^j\Big\}  =  \exp_\circ \Big\{\frac i\hbar \omega_{ij}a^i
          {(S^{-1})^j}_k y^k\Big\} \\
          &= U(S)\circ  \exp_\circ \Big\{\frac i\hbar \omega_{ij}a^i
          y^j\Big\} \circ U(S)^{-1}.
    \end{align*}
 \end{enumerate}

\end{proof}

\subsection{Representations}

 In an irreducible unitary representation of $H(V,\omega)$ the central elements $(0,t)$ act as
 scalars $e^{\frac i\hbar t}$. The Stone-von Neumann theorem (see
 e.g. \cite{folland:harmonic_analysis_in_ps}) states that up to unitary equivalence there is only one unitary
 representation of $H(V,\omega)$ for fixed $\hbar$.
 Such a representation $\rho$ gives rise to one of the Weyl algebra, where the
 action of the generators $y^i\in V^\ast$ is determined by
  $$ \rho( y^i)\psi  = -i\hbar\omega^{ij} \frac \partial {\partial
  t}\Big|_{t=0}\rho(t e_j,0)\psi,$$
 cf. (\ref{WeylOpsDefi}). The induced representation of
 Mp$(V,\omega)$ can be obtained more directly as follows. For $S\in
 $ Sp$(V,\omega)$ define another representation of $H(V,\omega)$ on the same Hilbert
 space through $$ \rho^S(v,t) = \rho(Sv,t).$$
 According to Stone-von Neumann $\rho$ and $\rho^S$ are unitarily
 equivalent, thus we can find $V(S)\in U(\mathcal H)$ s.t.
  \begin{equation}\label{MetaDefiRR}
        \rho(Sv,t) =V(S)\rho(v,t)V(S)^{-1}.
  \end{equation}
  Obviously $V(S)$ is not uniquely determined by $S$. We are going to show that it is unique up to a
 phase. Suppose that $\tilde V(S)$ satisfies equation \ref{MetaDefiRR}
 as well, then
  $$ [\tilde V(S)^{-1}V(S),W(a)]=0 \qquad \forall a\in V,$$
 and Schur's lemma implies $\tilde V(S)^{-1}V(S) \in \mathbb C\mathbf 1$. We have
  $$ V(S)V(T)W(a)V(T)^{-1}V(S)^{-1} =
  V(ST)W(a)V(ST)^{-1},$$
 and uniqueness of $V(ST)$ up to a phase implies
  \begin{equation}
    V(S)V(T)=c(S,T)V(ST),\qquad c(S,T)\in S^1.
  \end{equation}
 It is not possible to eliminate the phase $c(S,T)$ completely, but
 it can always be chosen to be $\pm 1$ \cite{folland:harmonic_analysis_in_ps},
 and comparison with equation \ref{MetaGroupDefi} of Lemma
 \ref{MetaProperties} leads us to conclude that $V(S)=\rho(U(S))$.

 The map Sym$(V^\ast)\rightarrow W(V^\ast)$ assigning to a polynomial on $V$ its
 element in the Weyl algebra is the quantization map on the flat
 phase space $(V,\omega)$. The essence of our discussion on the
 metaplectic and Heisenberg group is that the symmetry group $V\rtimes $ Sp$(V,\omega)$ of classical
 mechanics is replaced by its central extension $H(V,\omega) \rtimes
 $Mp$(V,\omega)$ in quantum mechanics. $\rtimes$ denotes the
 semidirect product, and Mp$(V,\omega)$ acts on $H(V,\omega)$ by the
 adjoint action: $U(S)\star W(a) =U(S)\circ W(a)\circ
 U(S)^{-1}=W(Sa)$.

\begin{bsp}[Schr\"odinger representation]
 We construct a representation of the Weyl algebra, following the
 well-known construction of spin representations of the Clifford
 algebra. Let $P$ be a Lagrangian subspace of $V$, meaning that
 its symplectic complement $P^\perp = \{ v\in V\ |\ \omega(v,p)=0 \ \forall p\in P\}$ equals
 $P$. Define $P^\flat =\{\omega(p,\cdot)\ |\ p\in P\} \subset
 V^\ast$. One can now introduce Darboux coordinates $q^i,p_j$ on
 $V^\ast$, such that $P^\flat$ is spanned by the $q^i$. The representation space is
 Sym$(P^\flat_\mathbb C)$, i.e. the space of complex polynomials in the $q^i$,
 and the action is determined by
  \begin{align}\label{FlatSchroedinger}
    \sigma(q)\psi &= q\psi \\
    \sigma(p) \psi &= [p,\psi]_\circ , \nonumber
  \end{align}
 for $q\in P^\flat_\mathbb C$, $p \in Q^\flat_\mathbb C:=$ span$\{p_1,\dots,p_n\}$, and
 $\psi\in $ Sym$(P^\flat_\mathbb C)$. Note that choosing different
 coordinates $p_j' = p_j + A _{jk}q^k$ does not change the
 representation because $[q^k,\psi]_\circ = 0$, so it really depends
 only on $P^\flat$, and not on the complementary subspace
 $Q^\flat$.

 (\ref{FlatSchroedinger}) is then extended to the whole Weyl algebra
 by requiring $\sigma$ to be a homomorphism w.r.t. the Weyl product
 (\ref{WeylProd}). Observe that for a basis element $p_i$ one
 obtains the well-known Schr\"odinger operator $\frac \hbar i
 \partial_{q^i}$.

 In our Darboux coordinates adapted to the decomposition $V=Q\oplus P$ the symplectic form
 is $ \omega= dp_j\wedge dq^j$. An element $X\in \mathfrak{sp}(V,\omega)$ can then be represented as
 $ X=\left(
              \begin{array}{cc}
                  A & B \\
                  C & -A^T \\
               \end{array}
      \right),$
 where $A,B,C \in \mathfrak{gl}(n,\mathbb R)$, $B^T=B,C^T=C$. The subset of
 $\mathfrak{sp}(V,\omega)$ respecting the polarization $P$ consists
 of matrices
  \begin{equation}\label{RealPolSymp}
        X=\left(
              \begin{array}{cc}
                  A & 0 \\
                  C & -A^T \\
               \end{array}
      \right).
  \end{equation}
 From the definition of $dU$ (\ref{SpMpAlgIsom}) we obtain
 the metaplectic representation in the Schr\"odinger picture:
  \begin{equation}\label{MetaSchroedinger}
        \sigma(dU( X))=\frac {i\hbar}2 B^{jk}\frac{\partial^2}{\partial
             q^j\partial q^k} - {A^k}_jq^j \frac
             \partial{\partial q^k} -\frac 1 2\text{tr }A +
             \frac i{2\hbar} C_{jk}q^jq^k .
  \end{equation}
\end{bsp}

\begin{bsp}[Fock-Bargmann representation]
 Let $P\subset V_\mathbb C$ be a Lagrangian subspace again, w.r.t.
 the complexified symplectic form, such that $P\cap \overline
 P=\{0\}$. Then we can introduce complex coordinates $z^j,\overline
 z^j$ on $V_\mathbb C^\ast$ such that $P^\flat$ is spanned by the
 $\overline z^j$, and $\omega$ assumes the complex standard form $i dz^j\wedge d\overline
 z^j$. The representation space is Sym$(\overline
 P^\flat)$, i.e. the space of holomorphic polynomials. The action of $W(V^\ast)$
 is determined by
  \begin{align}
     \beta(z)\psi&= z\psi, \\
     \beta(\overline z)\psi& = [\overline z,\psi]_\circ, \nonumber
  \end{align}
 leading to $\beta(\overline z^j)=\hbar \frac{\partial}{\partial z^j}$.
 Elements of the symplectic Lie algebra again take the form
  $ X= \left(
          \begin{array}{cc}
            A & B \\
            C & -A^T, \\
          \end{array}
        \right)$
 w.r.t. the splitting $V_\mathbb C = P\oplus \overline P$, where
 $A,B,C\in \mathfrak {gl}(n,\mathbb R)$, $B=B^T,C=C^T$. The ones
 respecting also the complex structure are
  \begin{equation}\label{ComplPolSymp}
        X = \left(
          \begin{array}{cc}
            A & 0 \\
            0 & -A^T, \\
          \end{array}
        \right),\qquad A\in \mathfrak{gl}(n,\mathbb R),
  \end{equation}
 and form a $\mathfrak u(n)$-algebra. The Fock space
 representation of the metaplectic algebra becomes
  \begin{equation}\label{MetaFock}
    \beta(dU(X)) = -\frac \hbar 2 B_{ab}\frac{\partial^2}{\partial
    z^a\partial z^b} - A_{ab} z^b\frac\partial{\partial z^a} -\frac
    12\text{tr }A + \frac 1{2\hbar} C_{ab}z^az^b.
  \end{equation}
\end{bsp}

 The representation spaces Sym$(\overline P^\flat)$ in these examples consist of polynomial
 functions on a subset of $V$. It is now possible to let $W(V^\ast)$
 act on more general functions, e.g. Schwartz functions $\mathcal
 S(Q)\subset L^2(Q)$ in the real case, $Q\cong\mathbb R^n$. We would
 like to find a suitable function space endowed with an inner
 product, such that the representations of $H(V,\omega)$ and Mp$(V,\omega)\subset W(V^\ast)$ become
 unitary. This will certainly be the case if the elements of $V^\ast$ act as self-adjoint
 operators on a common invariant, dense domain, which can be accomplished by
 the choice $L^2(Q,d^nq)$ in the real case, and $L^2_{hol}(V
 ,e^{-|z|^2/\hbar}d^nzd^n\overline z)$ in the complex one, where the index $hol$ denotes restriction
 to holomorphic functions. Is there a systematic way to find the
 inner product? The answer is yes, and the method is geometric
 quantization, where the inner product plays the role of a Hermitian
 structure on the prequantum bundle $B$, determined by the condition that a
 given connection on $B$ has to be metric. See section \ref{Sec_GeoQuant}.
 It is interesting to note that in the holomorphic case the Hilbert
 space is indeed spanned by polynomials in $z$, whereas in the real
 one it is not spanned by polynomials in $q$.

\subsection{Weyl bundle and metaplectic structures}

 On a symplectic manifold $(M,\omega)$ one has a symplectic vector space
 $(T_mM,\omega_m)$ over every point $m\in M$. One can then form the
 infinite-rank Weyl bundle $\mathcal W$, consisting of the collection of
 $W_m:=W(T^\ast_m M)$. Associated to $\mathcal W$ (or $TM$) is a principal
 $\mathbb R^{2n}\rtimes $ Sp$(2n)$ bundle (Sp$(2n)$ denotes Sp$(\mathbb R^{2n},dp_i\wedge dq^i)$),
 which can be lifted to a principal $H(2n)\rtimes$ Mp$(2n)$ bundle
 if the second Stiefel-Whitney class $w_2(M)$ vanishes. Then a unitary representation $\rho$ of
 $H(2n)$ induces one of $H(2n)\rtimes $ Mp$(2n)$ and gives rise to an associated Hilbert bundle $\mathcal H$ on
 $M$, such that $\mathcal W\subset $ End($\mathcal H)$.

\subsection{Analogy to the spin group}
 The metaplectic group is the symplectic analogue of the spin group.
 Let $(V,g)$ be a metric vector space, then its associated Clifford
 algebra Cl$(V^\ast)$ is, as a vector space, the antisymmetric tensor algebra $\Lambda
 V^\ast _\mathbb C$. The Clifford product is defined by
  $$ v \circ w = \frac 12v\wedge w + \frac 12g^{-1}(v,w),\quad \forall v,w\in
  V^\ast,$$
 which should be compared to the Weyl product (\ref{WeylProd}).
 The isomorphism from
  $$\mathfrak{so}(V,g)=\{A \in \text{End}(V)\ |\ g(Av,w) +
  g(v,Aw)=0\ \forall v,w\in V\}$$
 to the spin algebra is given by
  $$ dS: \mathfrak{so}(V,g)\rightarrow \mathfrak{spin}(V,g),\
  A\mapsto \frac 12 g_{ab}{A^b}_c \gamma^a\circ\gamma^c,$$
 where the $\gamma^i$ are basis elements of $V^\ast$, and thus
 satisfy $\{\gamma^i,\gamma^j\} :=\gamma^i\circ\gamma^j + \gamma
 ^j\circ \gamma ^i=g^{ij}$. The analogue of Lemma (\ref{MetaProperties})
 is given by
  $$    [dS(A),\gamma^i]_\circ = -{A^i}_j\gamma ^j $$
 and its exponentiated form. The representations of Cl$(V^\ast)$
 can be obtained in a manner similar to the Fock representation of $W(V^\ast)$

\section{Deformation Quantization}\label{sec_defQuant}
 We describe deformation quantization
 of symplectic manifolds according to Fedosov \cite{fedosov:deformationart,fedosov:deformationbook}. Let
 $(M,\omega)$ be a $2n$-dimensional symplectic manifold. Our goal is the
 construction of a star product $\ast$ on $C^\infty(M)$, i.e. a
 deformation of the ordinary point-wise product of functions on $M$.
  It turns out that we will have to consider $\hbar$ as a formal
 deformation parameter here. E.g. instead of $C^\infty(M)$ we have to deal with
  $$C^\infty(M)[[\hbar]] = \{ \sum_{k\geq 0} \hbar^ka_k\ |\ a_k \in
 C^\infty(M)\},$$
 where no sort of convergence property is imposed on the series $\sum_k
 \hbar^ka_k$. The results of section \ref{Section:MetaRep} will have
 to be interpreted in this sense as well then. In the
 following we will simply denote $C^\infty(M)[[\hbar]]$ by
 $C^\infty(M)$, and analogously for other structures.

 The following conditions are to be satisfied by $\ast$:
  \begin{enumerate}
    \item the coefficients $c_k$ of the product of
    $f=\sum_k\hbar^kf_k$ and $g=\sum_k\hbar^k g_k$:
        $$ f \ast g = \sum_{k=0}^\infty
        \hbar^kc_k(f,g)$$
      are bi-differential operators (of finite order).
    \item $c_0(x)=f_0(x)g_0(x)$
    \item the correspondence principle
      $$ [f,g]_\ast =f\ast g-g\ast f = i\hbar \{f_0,g_0\} +
      O(\hbar^2), $$
     holds, where $\{f,g\}=\omega^{-1}(df,dg)=\omega^{ab}\partial_a f\partial_bg$ denotes the poisson bracket defined by
     $\omega$.
  \end{enumerate}

\subsection{The Fedosov connection}

 In every point $m\in M$ we can form the Weyl algebra $W_m =
 W(T^\ast_m M)$, whose elements have the form
  $$\sum_{k,r\geq 0, } \hbar^k f_{k;l_1,\dots,l_r} y^{l_1}\dots
  y^{l_r}$$
 (including a sum over the $l_i$), with multiplication determined by
 (\ref{WeylProd}). The collection of local Weyl algebras $W_m$ forms
 the Weyl bundle $\mathcal W$, whereas in the flat case $M=V$ there was only one Weyl
 algebra and representation. We want to get rid of these extra
 algebras by identifying 'neighboring' Weyl algebras, which can be done by
 introducing a flat connection on $\mathcal W$. Choose
 a symplectic connection $\nabla$ on $TM$, i.e. one satisfying
 $\nabla \omega=0$ for the induced connection on $T^\ast M\otimes T^\ast M$,
 with vanishing torsion. Symplectic connections always exist, but contrary to the Riemannian case they are not unique.
 It induces a connection on the whole tensor
 algebra of $M$, in particular one on $\mathcal W= $ Sym$(T^\ast
 M)$, through the Leibniz formula. In local Darboux coordinates
 $\nabla$ can be decomposed as $\nabla=d+ \Gamma$, where $\Gamma\in
 \Omega^1\big(U;\mathfrak{sp}(2n)\big)$ is a Lie algebra valued
 1-from. Its components are defined by $\Gamma^k_{ij}e_k =
 \Gamma(e_i)\cdot e_j$, in the local Darboux basis $\{e_i\}$ of
 $TM$. Further we define $\Gamma_{ijk}=\omega_{il}\Gamma^l_{jk}$,
 which is symmetric in all indices due to $\nabla$ being
 torsion-free and symplectic. The curvature of $\nabla$ is defined
 as $R=\nabla^2 \in \Omega^2(M;T^\ast M\otimes TM)$, with components
  \begin{equation}\label{curvatureFromConnection}
        {R^i}_{jkl} =
        \partial_k\Gamma^i_{lj}-\partial_l\Gamma^i_{kj} +
        \Gamma^i_{km} \Gamma ^m_{lj} - \Gamma^i_{lm} \Gamma ^m_{kj},
  \end{equation}
 and $R_{ijkl}=\omega_{im}{R^m}_{jkl}$ is symmetric in the first two
 (Lie algebra) indices, and antisymmetric in the last two
 (differential form) indices. We denote the local basis elements of $T_m^\ast
 M$ by $y^i$ when considered as elements of the Weyl algebra $W_m$,
 and by $dx^i$ when considered as elements of the
 exterior algebra $\Lambda T^\ast_m M$. Differential forms with
 values in the Weyl bundle are sections of $\Omega(\mathcal
 W)=\Gamma(\mathcal W \otimes \Lambda T^\ast M).$ Note that, as a
 vector bundle, $\mathcal W \otimes \Lambda T^\ast M$ is just the
 full tensor bundle of $M$. On $\Omega(\mathcal W)$ we write $\circ$ for
 $\circ \otimes \wedge$, and use the graded commutator
  $$ [\xi,\eta] = \xi\circ \eta - (-1)^{pq} \eta\circ \xi,$$
 for $\xi,\eta$ differential forms of degree $p,q$ respectively,
 with values in $\mathcal W$.
 The connection is extended to $\Omega(\mathcal W)$ through
  \begin{align*}
    \nabla (\xi\circ \eta) &=\nabla \xi\circ\eta + (-1)^q\xi
    \circ\nabla\eta, \qquad \xi \in\Gamma(\mathcal W\otimes\Lambda^q) \\
    \nabla (\phi\wedge \eta ) &=d\phi \wedge \eta + (-1)^q \phi\wedge
     \nabla \eta ,\qquad \phi\in \Omega^q(M).
  \end{align*}

  Explicitly we have
  $$ \nabla y^{i_1}\dots y^{i_k} = -\sum_j \Gamma^{i_j}_{ab}
  y^{i_1}\dots \breve y^{i_j} \dots y^{i_k}y^a dx^b,$$
 where $\breve y^{i_j}$ means omitting the element, and which can be written in the form
  \begin{equation}
    \nabla  = d+ [dU(\Gamma),\cdot] = d-\frac i{2\hbar}
    \Gamma_{ijk}[y^iy^j,\cdot]dx^k.
  \end{equation}
 Here $dU$ is the isomorphism between the symplectic and
 metaplectic Lie algebras (\ref{SpMpAlgIsom}).
 Now we introduce two further operators on $\Omega(\mathcal W)$:
  \begin{equation}
    \delta = dx^k\wedge \frac{\partial}{\partial y^k},\qquad
    \delta^\ast= y^k \iota\Big(\frac{\partial}{\partial x^k}\Big),
  \end{equation}
 where $y^k$ denotes multiplication with $y^k$ (w.r.t. the symmetric tensor product), and
 the contraction $\iota\Big(\frac{\partial}{\partial x^k}\Big)$ acts
 only on the form part. In brief, $\delta$ replaces one of the $y^i$ by $dx^i$,
 whereas $\delta^\ast$ replaces $dx^i$ by $y^i$. One easily checks
\begin{lem}\
 \begin{enumerate}
   \item $\delta^2=(\delta^\ast)^2=0$
   \item Applied to $ y^{i_1}\dots y^{i_l} dx^{j_1}\wedge\dots \wedge dx^{j_p}
   $ the following identity holds:
     \begin{equation}
        \delta \delta^\ast+ \delta^\ast \delta =( l + p)id.
     \end{equation}
 \end{enumerate}
\end{lem}
 We also define $\delta^{-1}$ by
 \begin{equation}
        \delta^{-1} = \frac 1{l+p}\delta^\ast
 \end{equation}
 for $l+p>0$, and $\delta^{-1}=0$ otherwise. If the projection of an
 element $\xi \in\Omega(\mathcal W)$ to its part in
 $C^\infty(M)\subset\Omega(\mathcal W)$ is denoted by $\xi_{00}$, then
 the following decomposition holds, analogously to the Hodge-de
 Rahm decomposition of forms:
  \begin{equation}\label{decompFed}
    \xi = \delta\delta^{-1} \xi + \delta^{-1}\delta \xi + \xi_{00}.
  \end{equation}
\begin{defi}
 The $\hbar$-degree of a homogeneous element
  \begin{equation}
    \sum_m \hbar^k \xi_{k; i_1,\dots,i_l,j_1,\dots,j_m}
   y^{i_1} \dots y^{i_l}dx^{j_1}\wedge \dots\wedge dx^{j_m} \quad
   \in\ \Omega(\mathcal W)
  \end{equation}
 (no sum over $k,l$) is defined as $k+l/2$.
\end{defi}
 From its definition (\ref{WeylProd}) it follows that $\circ$ respects the
 gradation on $\mathcal W$ defined by the $\hbar$-degree. The subspaces of
 homogeneous elements of $\hbar$-degree $j$ are denoted by $\mathcal
 W_j$. Now $\mathcal W$ carries two gradations, the other one being defined by the degree in
 Sym$(T^\ast_\mathbb C M)$, which is respected by the symmetric tensor
 product, but not by $\circ$. The corresponding projections are
  \begin{equation}\label{GradProjektionen}
        \pi_\hbar^j :\mathcal W \rightarrow \mathcal W_j,\qquad \text{and}\qquad
         \pi_\otimes^k: \mathcal  W \rightarrow \text{Sym}_k(T^\ast_\mathbb C
         M),
  \end{equation}
and we also adopt the convention to denote $\pi_\otimes^0(\xi)$ by
 $\xi_0$. It is important to note that $\delta$
 decreases the $\hbar$-degree by $1/2$, whereas $\delta^{\ast}$
 increases it.

 Now we come to the construction of a flat connection
 on $\mathcal W$, with curvature $\Omega=\frac
 i\hbar[\omega,\cdot]=0$. The ansatz
  \begin{equation}\label{Connection}
    D=\nabla +\frac i\hbar[\gamma,\cdot]= d+ [dU(\Gamma)+
    \frac i\hbar\gamma,\cdot],
  \end{equation}
 with an as yet undetermined 1-form $\gamma\in \Omega^1(\mathcal
 W)$, leads to the curvature $\Omega=D^2=\frac i\hbar[\tilde \Omega,\cdot]$,
 where
  \begin{equation}
   \tilde \Omega=\hat R+ \nabla \gamma + \frac
  i\hbar\gamma^2.
  \end{equation}
 Here $R$ denotes the curvature 2-form of $\nabla$, $\hat R=\frac \hbar idU(R) =-\frac
 1{4} R_{ijkl}y^iy^jdx^k\wedge dx^l$, and $\gamma^2=\gamma\circ\gamma$. Of course, $\gamma$ is
 only determined up to addition of a scalar form. We require the normalization
 $\gamma_0=0.$ In the flat case
 $M=\mathbb R^{2n}$ with standard symplectic form $\omega=dp_i\wedge dq^i$ we can choose
 $\Gamma=0$, and $\gamma=\omega_{ab}y^bdx^a$ leads to
 $\tilde\Omega= \omega$. Therefore, in the general case we
 split $\gamma$ as
  \begin{equation}
   \gamma=\omega_{ab}y^bdx^a + r.
  \end{equation}
 Observing that $\delta$ can be written in the form $\delta \xi =
 -\frac i\hbar \omega_{ab}dx^a[y^b,\xi]$, we obtain for the curvature
    $$  \tilde\Omega= \omega +\hat R- \delta r
          +\nabla r +\frac i\hbar r^2,$$
 so that $\tilde \Omega= \omega$ becomes equivalent to
  \begin{equation}\label{abelscherZshgBedg}
    \delta r = \hat R +\nabla r + \frac
     i\hbar r^2.
  \end{equation}

 \begin{theo}[Fedosov]
  Under the condition $\delta^{-1}r=0$, eq.
  (\ref{abelscherZshgBedg}) has exactly one solution $r$.
 \end{theo}
\begin{proof}[Sketch of proof]
 For a 1-form we have $r_{00}=0$, which together with the condition
 $\delta^{-1}r=0$ implies that the decomposition (\ref{decompFed}) takes
 the form $r=\delta^{-1}\delta r$. Applying $\delta^{-1}$ to
 (\ref{abelscherZshgBedg}) leads to
   \begin{equation}\label{abelscherZshgBedg2}
     r = \delta^{-1} \hat R +\delta^{-1} \Big(\nabla r +
     \frac i\hbar r^2\Big).
   \end{equation}
 As $\nabla$ preserves the $\hbar$-filtration, whereas $\delta^{-1}$
 increases the degree by 1/2, it follows by iteration that
 (\ref{abelscherZshgBedg2}) has exactly one solution. The iteration steps are
  \begin{equation}\label{FedConnIteration}
    r^{(3)}=\delta^{-1}\hat R,\quad r^{(n+1)}=\delta^{-1}\hat R
    +\delta^{-1}\big(\nabla r^{(n)}+\frac i\hbar (r^{(n)})^2\big) \mod \hbar^{n/2+1},
  \end{equation}
 and $r=\lim_{n\rightarrow \infty} r^{(n)}$. Here$\mod \hbar^{n/2+1}$
 means discarding all terms of $\hbar$-degree at least $n/2+1$.
 These terms are not stable under iteration yet, and it is
 convenient, although not necessary, to ignore them. The condition
 $\delta^{-1}r=0$ is fulfilled due to $(\delta^{-1})^2=0$. For the
 proof that $r$ indeed solves (\ref{abelscherZshgBedg}) we refer to
 Fedosovs texts \cite{fedosov:deformationart,fedosov:deformationbook}.
\end{proof}
 The first iteration gives
 \begin{equation}\label{curvIterations}
   r^{(4)} = -\frac 18 R_{ijkl}y^iy^jy^k dx^l -\frac 1{40}\tilde\nabla_m R_{ijkl}y^iy^jy^ky^mdx^l
 \end{equation}

 where $\tilde\nabla$ is the product connection on $\mathcal
 W\otimes \Lambda$, thus acts the same way on $y^i$ and $dx^i$.

\subsection{Observables and star product}

 Having constructed a flat connection on $\mathcal
 W$ we want to identify the quantum operators with the set of flat sections of
 $\mathcal W$ with respect to $D$, i.e. those satisfying $D\hat
 f=0$. As $D=\nabla- \delta+\frac i\hbar [r,\cdot]$, this equation can be written
 in the form
  \begin{equation}\label{flacheObservBedg}
        \delta \hat f=\nabla \hat f + \frac i\hbar [r,\hat f].
  \end{equation}
 We denote the set of flat (or parallel) sections by $\Gamma_D(\mathcal W)$.
\begin{theo}[Fedosov]
 To every $f\in C^\infty(M)$ there exists exactly one $\hat
 f\in \Gamma_D(\mathcal W)$ such that $\hat f_0=f$.
\end{theo}
\begin{proof}[Sketch of proof]
 For a 0-form $\hat f$ we have $\delta^{-1}\hat f=0$ and $\hat f_{00}=\hat f_0$,
 so that the decomposition (\ref{decompFed}) becomes $\hat f=\hat f_0 + \delta^{-1}\delta \hat
 f$. Then equation (\ref{flacheObservBedg}) implies
  \begin{equation}
   \hat f=\hat f_0+\delta^{-1}\big(\nabla \hat f+\frac i\hbar[r,\hat
   f] \big).
  \end{equation}
 Again this equation has a unique solution, which can be determined by
 iteration:
  \begin{equation}\label{OperatorIterationGlg}
   \hat f^{(0)}=\hat f_0=f,\quad \hat f^{(n+1)}=\hat
   f_{0}+\delta^{-1} \big(\nabla \hat f^{(n)}+\frac i\hbar [r,\hat
   f^{(n)}]\big) \mod \hbar^{n/2+1}.
  \end{equation}
 For the proof that $\hat f$ indeed solves (\ref{flacheObservBedg}) we
 again refer to \cite{fedosov:deformationart,fedosov:deformationbook}.
\end{proof}

 The first iterations give:
 \begin{align}\label{OperatorIteration}
    \hat f^{(3)} &= f+y^k\nabla_kf + \frac 12y^j\nabla_jy^k\nabla_k
       f + \frac 16 y^i\nabla_iy^j\nabla_j y^k\nabla_k f \nonumber\\
       &\qquad\qquad +\frac 1{24} R_{abcd}\omega^{ck}\partial_k f y^ay^by^d
 \end{align}

 An explicit calculation of the covariant derivatives shows that in
 third order one has
 \begin{align}\label{OperatorIteration2}
    \hat f^{(3)}&=f+\partial_kfy^k + \frac
       12\omega_{jl} (\nabla_k X_f)^l y^jy^k + \\
      &\qquad+ \frac 16\big[\omega_{jl}(\nabla_i\nabla_kX_f)^l +
      \Gamma_{ijl}(\nabla_kX_f)^l-\frac
      14R_{ijkl}X_f^l\big]y^iy^jy^k, \nonumber
 \end{align}
 where $X_f$ is the Hamiltonian vector field of $f$, defined by
 $X_f(g)=\omega^{-1}(dg,df)$. Note that a solution of the equation
 $D\hat f=0$ is not uniquely determined by its value at a point
 $m\in M$. This value is determined by the Taylor series of $f$ at
 $m$, which does not fix $f$.

 The quantization map
  $$ (\pi_\otimes^0)^{-1} : C^\infty(M) \rightarrow
  \Gamma_D(\mathcal W),\ f\mapsto \hat f$$
 allows for the definition of a star product on
 $C^\infty(M)$, namely
  \begin{equation}\label{StarProdDefi}
         f\ast g:= \pi_\otimes^0(\hat f\circ \hat g).
  \end{equation}
 Using the explicit expression (\ref{OperatorIteration2}), as well
 as the relation
  \begin{equation}\label{DefoQuantContraction}
        \pi_\otimes^0(y^{a_1}y^{a_2} \dots y^{a_k} \circ y^{i_1}y^{i_2}\dots y^{i_k}) =
  \Big( \frac {i\hbar}2\Big)^k \sum_{\pi \in S_k} \omega^{a_1
  i_{\pi(1)}} \dots \omega^{a_k i_{\pi(k)}},
  \end{equation}
 we arrive at
  \begin{equation}\label{StarProdExpl}
          f\ast g = fg - \frac {i\hbar}2 \omega(X_f,X_g) + \frac{\hbar^2}4
      (\nabla_jX_f)^b(\nabla_bX_g)^j+ O(\hbar^3).
  \end{equation}
 The conditions for a star product mentioned at the beginning of the
 section are easily checked, using the fact that the Poisson bracket can be expressed as
 $\{f,g\}= \omega(X_g,X_f)$. For $M=\mathbb R^{2n}$ with $\Gamma=0$ one obtains
  \begin{equation}\label{vollerOperatorflach}
    \hat f = \sum_{k=0}^\infty \frac 1{k!}(\partial_{i_1} \dots
    \partial_{i_k} f) y^{i_1}\dots y^{i_k},
  \end{equation}
 and thus the Groenewold-Moyal product \cite{Groenewold_QM,
 moyal:original}
  \begin{equation}
   f\ast g(x)=\exp\Big(\frac {i\hbar}2\omega^{ij}\frac \partial{\partial
   x^i}\frac \partial{\partial y^j}\Big) f(x)g(y)\bigg|_{y=x},
  \end{equation}
 or $f\ast g = \mu\circ\exp\big\{\frac {i\hbar}2 \omega^{-1}\big\} f\otimes g$,
 recalling that $\omega^{-1}\in \Gamma(TM\otimes TM)$ acts naturally on
 $C^\infty(M)\otimes C^\infty(M)$; $\mu (f\otimes g)(x):=f(x)g(x)$.

\bigskip

 If $M$ admits a metaplectic structure $\mathcal H\rightarrow M$,
 then we know that $\mathcal W \subset $ End$(\mathcal H)$. Now
 $\Gamma_D(\mathcal W)$ is our algebra of quantum
 observables, and it is natural to conjecture that the physical states
 are sections of $\mathcal H$. But again $\Gamma(\mathcal H)$ is too
 large, and we need a way to identify different fibres of $\mathcal
 H$. The most obvious way to do this would be by means of a flat
 connection, as in the case of the observables \cite{reuter_QMGaugetheory}.
 The connection form of $ D$ on $\mathcal W$ is given by
 a commutator, i.e. $D = d+[A,\cdot]$, where
  \begin{equation}\label{FedconnectionForm}
        A=dU(\Gamma)+\frac i\hbar \gamma = \frac i\hbar \omega_{ij}y^jdx^i  -\frac i{2\hbar}
        \omega_{ij}\Gamma^j_{kl}y^iy^l dx^k  -\frac i{8\hbar} R_{ijkl}y^iy^jy^k
        dx^l + \dots,
  \end{equation}
from (\ref{Connection}), implying that $D$ is the induced connection
of $D_\mathcal H=d+A$ on $\mathcal H$. However, $D_\mathcal H$ is
not flat, as $D^2_\mathcal H= \frac i\hbar \omega\neq 0$, and
therefore $\Gamma_D(\mathcal H)=\{\psi\in \Gamma(\mathcal H)\ |\
D_\mathcal H\psi=0\}$ consists only of the zero section. We will
find out how to get a flat connection in section
\ref{Sec:FullQuant}.

\section{Geometric Quantization}\label{Sec_GeoQuant}

 Geometric quantization is a prescription for constructing a Hilbert
 space ${\mathcal H_P}$ (the space of states) on a symplectic manifold, as well as a
 few quantum operators acting on it. Our main reference for the material presented here
 is Woodhouse's book \cite{woodhouse:quantisierung}.
\subsection{The Hilbert space}
 ${\mathcal H_P}$ is not uniquely
 determined by $(M,\omega)$ but depends on further structure:
 \begin{itemize}
   \item a 'prequantum bundle' $B\rightarrow M$, i.e. a Hermitian
   line bundle over $M$, with metric connection $\nabla^B$ of curvature $-\frac i\hbar \omega$.
   \item a metaplectic structure $\mathcal H\rightarrow M$.
   \item a polarization, i.e. a strongly integrable (\cite{woodhouse:quantisierung}, p.92)
    Lagrangian subbundle $P\subset T_\mathbb CM$, Lagrangian meaning that
    $$ P_m^\perp:= \{v\in {T_m}_\mathbb CM\ |\ \omega(v,p)=0 \ \forall p\in
    P_m\}$$
   equals $P_m$. In particular this implies rk$_\mathbb CP=\frac 12$dim$_\mathbb RM$.
 \end{itemize}

 Existence of $B$ implies the condition $[\omega/2\pi\hbar] \in
 H^2(M,\mathbb Z)$, whereas the metaplectic structure exists iff the
 second Stiefel-Whitney class $w_2(M)$ of $M$ vanishes. We will
 assume these conditions to be satisfied, although they can be
 slightly weakened for pure geometric quantization, see
 \cite{rawnsley:metaplectic}, or \cite{woodhouse:quantisierung},
 p. 233. The existence problem for polarizations is quite subtle, and
 not completely solved as it seems. In particular there are symplectic manifolds that do not
 admit any polarization at all \cite{gotay_NonexistencePol}.

 Different prequantum bundles correspond to distinct physical situations,
 e.g. different magnetic monopole charges, or
 different vacuum $\theta$-angles \cite{Nair_QFT-AMP}. The same might be true
 for the metaplectic structure, but it is not obvious how to
 interpret the polarization; it is usually claimed that
 'most' polarizations can be considered as equivalent, but this
 seems not to be possible for all of them, and is definitely not in the
 infinite-dimensional case, as demonstrated by Shale's theorem
 \cite{shale_shaleTheo,robinson_infiniteMeta}.

 Suppose a choice of the above mentioned structures $(B\rightarrow M,\mathcal H\rightarrow M,P )$
 has been made. The polarization can be transferred to the
 cotangent bundle through
  $$ \flat: T_\mathbb CM \rightarrow T^\ast_\mathbb C M,\ X^\flat(Y)
  =\omega(Y,X) = \omega_{ab} Y^aX^b.$$

\begin{bsp}[Real Polarizations]
 Suppose $P$ is the complexification of a subbundle of
 $TM$. In this case we have $P\cap\overline P=P$, the polarization is called real, and one can
 locally find Darboux coordinates $(q^1,\dots,q^n,p_1,\dots,p_n)$ on $M$ (i.e. $\omega=dp_i\wedge dq^i$) such that
  $$P=\text{span}\{\partial_{p_i}\ |\ i=1,\dots,n\},\quad P^\flat=\text{span}\{ dq^j\ |\
  j=1,\dots,n\}.$$
 These coordinates are said to be adapted to $P$.
 Define $Q_m=$ span$\{\partial_{q^i}\ |\ i=1,\dots,n\}$, then
 the local Hilbert space $\mathcal H_m$ from the metaplectic structure is the Schr\"odinger
 representation space $L^2(Q_m)$ on which $dq^j|_m, dp_i|_m\in W_m$
 act as the standard Schr\"odinger operators $\hat q^j,\hat p_i$.

\end{bsp}
\begin{bsp}[Complex Polarizations]
 Suppose $P\cap P =\{0\}$, or equivalently $P\oplus \overline P=T_\mathbb CM$. In this case the polarization is called
 totally complex, and defines a complex structure $J$ on $M$ in the
 following way. Split $X\in TM$ as $X=Z+\overline Z$, where $Z\in
 P$, and define $JX=iZ-i\overline Z$.

 We assume that the associated symmetric bilinear form
 $g(X,Y):=\omega(X,JY)$ is positive definite, and therefore gives
 $(M,\omega)$ the structure of a K\"ahler manifold. One can then
 choose complex coordinates $z^a$ on $M$, s.t.
  $$ P = \text{span}\{ \partial_{z^a}\ |\ a=1,\dots,n\},\quad P^\flat
   =\text{span}\{d\overline z^a\ |\ a=1,\dots,n\},$$
 and $\omega$ assumes the standard form
   $ \omega=  i \partial\overline \partial K$
 for some real valued function $K\in C^\infty(M)$, the so called
 Kähler potential. If $\omega_m= iK_{a\overline b}dz^a \wedge d\overline z^b|_m $, then
 the local Hilbert space $\mathcal H_m$ is the Fock space $L^2_{hol}(T_{m
 }M,\ e^{-K_{a\overline b}w^a\overline w^b/\hbar}dwd\overline w)$, where $w^a$ are
 the induced coordinates on $T_{m}M$. $dz^a|_m$ and $d\overline z^b|_m$
 act on $\mathcal H_m$ as multiplication with ${w}^a$ and differentiation $\hat {\overline w}^b=\hbar K^{a\overline b}
 \frac\partial{\partial {w}^a}$ respectively, where as usual
 $K^{a\overline b}$ are the matrix elements of the inverse matrix.
 In particular we have $(\hat w^a)^\dagger = \hat{\overline w}^a$.

\end{bsp}

 The so called canonical line bundle $K^P$ on $M$ associated to $P$ has
 fibre
  \begin{equation}
   K^P_m =\{\alpha \in \Lambda^n T^\ast_{m\mathbb C} M\ |\ \iota_X\alpha=0\ \forall
   X\in\overline P_m\},
  \end{equation}
 where $\iota_X$ denotes contraction with $X$. In the real case
 $K^P$ is spanned (locally) by $d^nq$, in the complex one by $d^n
 z$. Its definition requires a reduction of the structure group
 Sp$(2n)$ of $M$ to the subgroup preserving the polarization. For real
 polarizations the corresponding Lie algebra elements have the form
 (\ref{RealPolSymp}), where $A$ can be restricted to
 $\mathfrak{so}(n)$, as it is always possible to reduce to a maximal
 compact subgroup. Then $K^P$ has structure group $\mathbb Z_2$, which we assume to be
 further reducible to the trivial group. In the complex case the
 Lie algebra of the structure group is $\mathfrak u(n)$, spanned by elements
 (\ref{ComplPolSymp}), and $K^P$ has structure group $U(1)$.

 Further we need a square root of $K^P$, i.e. a line bundle $\Delta^P$ such
 that $\Delta^P\otimes \Delta^P = K^P$. Such a square root exists iff $M$
 admits a metaplectic structure; we will describe how this comes
 about at the end of this section. Elements of $\Delta^P$ will be denoted by
 square roots of $n$-forms. In the real case $\Delta^P$ is spanned
 locally by $\sqrt{d^nq}$, in the complex one by $\sqrt{d^nz}$.

   Now we introduce the partial connection on $\Delta^P$, which
 only allows differentiation in $\overline P$-direction. It is just the exterior derivative
 restricted to $\overline P$-vectors:
  \begin{equation}
         \nabla_{\overline X} \nu =  \overline X(\nu) \qquad
  X\in \Gamma(P),\nu\in\Gamma(\Delta^P).
  \end{equation}
 That the r.h.s. is well-defined can be seen from the fact that
 $\nabla$ is induced by a partial connection on $K_P$, which
 coincides with the exterior derivative acting on $\Omega^n(M)$
 there. Consult \cite{woodhouse:quantisierung} for details.

\begin{defi}\label{PolarizedDefi}
 Let $E\rightarrow M$ be a vector bundle with (partial) connection.
 A section $\psi$ of $E$ is called polarized if $\nabla_{\overline
 X}\psi=0$ for any $X\in \Gamma(P)$, and the set of polarized
 sections is denoted $\Gamma_P(E)$. If we want to emphasize the
 dependence on $\nabla$ we also write $\Gamma_{P;\nabla}(E)$. Further we define polarized
 functions to be elements of
  $$ C^\infty_P(M)=\{f\in C^\infty(M)\ |\ \overline X(f)=0
  \ \forall X\in \Gamma(P)\}.$$
\end{defi}

 $\Gamma_P(E)$ is a $C_P^\infty(M)$-module. Let $P,P'$ be two
 polarizations, and $E,F$ vector bundles with connections. Then one can take the tensor product
  $\Gamma_P(E)\otimes \Gamma_{P'}(F)$ in the category of
  $C^\infty_{P\oplus P'}(M)$-modules. However, the resulting space can be
  made into a $C^\infty_{P\cap P'}(M)$ module, which we also denote by the tensor
  product. One easily checks that
 \begin{equation}\label{polTensorProd}
  \Gamma_P(E)\otimes \Gamma_{P'}(F) \subset \Gamma_{P\cap P'}(E\otimes
  F).
 \end{equation}

 We only cite here the result that there is an isomorphism
 $\Delta^P\otimes\overline  \Delta^P\cong\pi^\ast\Delta(M/D)$, where
 $D=(P\cap \overline P)\cap TM$, the space of maximal connected
 integral manifolds of $D$ is denoted by $M/D$, and $\Delta(M/D)$ is
 the density bundle on $M/D$. One should think of its sections (called densities) as absolute
 values of differential forms of top degree, which can be integrated
 over $M/D$. $\pi: M \rightarrow M/D$ is the projection. We assume here and in the following
 that $D$ has constant rank, and $M/D$ is an orientable manifold. In the real case $D$ has rank
 $n$, for totally complex polarizations $D$ is trivial and $M/D=M$. The isomorphism of line bundles induces one
 on the level of sections:
  \begin{equation}\label{densityPairing}
        \Gamma_{P\cap \overline P}(\Delta^P\otimes \overline \Delta^P) \rightarrow \Gamma\big(\Delta(M/D)\big),
  \end{equation}
 which, due to (\ref{polTensorProd}), gives rise to a sesquilinear pairing
  $$ (\cdot,\cdot) : \Gamma_P(\Delta^P)\times \Gamma_{P}(\Delta^P)
  \rightarrow \Gamma\big(\Delta(M/D)\big).$$
 The restriction to polarized sections is needed to ensure that the
 resulting section of $\pi^\ast \Delta(M/D)$ is constant on the integral manifolds of $D$, and so
 defines a section of $\Delta(M/D)$.
 In the real case we have $(\sqrt{d^nq} ,\sqrt{d^nq})=|d^nq|$, in the complex
 situation we get $(\sqrt{d^nz},\sqrt{d^nz})=(\det g)^{1/4}d^nz\wedge
 d^n\overline z$, where the factor $(\det g)^{1/4}$ comes from the
 canonical trivialization of $\Delta^P\otimes \overline \Delta^P$,
 determined by the global, nonvanishing section $(\det g)^{1/4}\sqrt
 {d^nz} \otimes \sqrt{d^n\overline z}$. For the general case see \cite{woodhouse:quantisierung},
 p.230-234.

 On $ \Delta^P\otimes B$ we have the product partial connection, and
 consider the set of polarized sections $\Gamma_P(\Delta^P\otimes B)$.
 It carries a pairing $\langle\cdot,\cdot{\rangle_{}}_{G}$, given by
  \begin{equation}\label{pairingGQ}
        \langle  \mu\otimes a, \nu\otimes b {\rangle_{}}_{G} = \int_{M/D} (\mu,\nu) \langle
  a,b\rangle_B .
  \end{equation}
 Our quantum Hilbert space is
 then of course the space of square-integrable elements
  \begin{equation}
    \mathcal H_P = L^2_P(\Delta^P\otimes B):= \{\psi \in
    \Gamma_P(\Delta^P\otimes B)\ |\ \langle \psi,\psi{\rangle_{}}_{G} <\infty\}.
  \end{equation}
\subsection{Observables}
 We do not give the general construction of the observables here.
 Suffice it to say that only a very limited class of functions on
 $M$ can be quantized in the traditional formalism of geometric
 quantization, in the real case these are the functions which are
 affine-linear in the momentum: $f(q,p) = \alpha(q) +
 \beta^i(q)p_i$. The corresponding quantum observable is an operator on
 $\mathcal H_P$ (unbounded in general). We can choose a local trivialization
  in which the connection on $B$ takes the form $\nabla^B = d-\frac i\hbar \theta$,
 where $\theta=p_idq^i$. Wave functions can then be represented as
 $\psi \otimes \sqrt{d^nq}$, where $\psi$ is a function of $q$ alone,
 the operators are the canonical ones
  \begin{equation}\label{GeomQuantOps}
    \hat q^j (a \otimes \sqrt{d^nq} )(q)= q^j  a(q) \otimes \sqrt{d^nq} ,\qquad
    \hat p_j(a \otimes \sqrt{d^nq} )(q) = -i\hbar
    \frac\partial{\partial q^j}a(q) \otimes \sqrt{d^nq},
  \end{equation}
 and the function $\beta^i(q)p_i$ is mapped to the symmetrized
 product of the operators $\beta^i(\hat q)$ and $\hat p_i$.

\bigskip

 On a K\"ahler manifold we have $\omega= i\partial\overline \partial
 K$, a symplectic potential is $\theta=-i\partial K$, $B$ is a holomorphic line bundle, and wave functions
 take the form $\psi(z)\otimes \sqrt{d^nz}$. The Hermitian metric on
 $ B$ is nontrivial however, the condition for the connection
 $\nabla^B$ to be compatible with $\langle\cdot,\cdot\rangle_B$
 implies
  \begin{equation} \label{KaehlerPreQMtr}
        \langle \psi,\phi\rangle_B = \psi(z)\overline{\phi(z)} e^{-\frac
  K{\hbar}}.
  \end{equation}
 Holomorphic functions act as multiplication operators, and
 $\partial_a K:=\frac{\partial K}{\partial z^a}$ as
  \begin{equation}\label{GeomQuantKaehlerOp}
     \widehat {\partial_a K}\big( \psi \otimes
    \sqrt{d^n z}\big)(z)= \hbar \frac{\partial}{\partial z^a} \psi(z) \otimes \sqrt{d^n
    z}.
  \end{equation}

\subsection{The role of $\mathcal H$}
 The line bundle $\Delta^P$ is closely related to the bundle
 $\mathcal H$ of symplectic spinors \cite{kostant_symplecticspinors}. The situation turns out however to be
 essentially different for real and totally complex polarizations.

\paragraph{Real polarizations}
 In this case $(\Delta^P)^{-1}$ can be identified with the subbundle
 \begin{equation}\label{KostantsConstr}
   \{\psi\in \mathcal H\ |\ y \psi=0\ \forall y\in
   P^\flat\} \subset \mathcal H,
 \end{equation}
 where we consider $y\in T^\ast M$ as an element of the Weyl algebra
 $\mathcal W\subset $ End$(\mathcal H)$.
 The space $P^\flat$ is spanned by the $\hat q^j$, $\mathcal H_{m}$ is the Schr\"odinger representation
 space $L^2(Q_m)$, where $Q_m\subset T_mM$ is spanned by the
 $\frac\partial{\partial q^j}$, and $(\Delta^P)^{-1}$ is generated by the
 $\delta$-function. The latter transforms as
  $$ \delta \mapsto \frac 12 \text{tr}(A)\delta$$
 under a symplectic gauge transformation of $TM$ of the form (\ref{RealPolSymp}),
 which can be seen from (\ref{MetaSchroedinger}). Note that this is the right
 transformation behavior for $(\Delta^P)^{-1}$.

\paragraph{Kähler polarizations}
 In the complex case $\mathcal H_m$ is the Fock representation space
 consisting of holomorphic functions on $T_mM$, and $P^\flat$
 is spanned by the $\hat {\overline z}^j$, which act as
 differentiation operators. Therefore the (vacuum) line bundle
 \begin{equation}\label{KostantsConstrCplx}
   \{\psi\in \mathcal H\ |\ y \psi=0\ \forall y\in
   P^\flat\},
 \end{equation}
 is spanned by the constant functions on $T_mM$, which transform under an (infinitesimal) symplectic transformation
 of the form (\ref{ComplPolSymp}) as
 \begin{equation*}
   c \mapsto -\frac 12\text{tr}(A) c,
 \end{equation*}
 according to (\ref{MetaFock}). But this is the transformation behavior
 of $\Delta^P$ itself, which we therefore identify with
 (\ref{KostantsConstrCplx}). \\
 Note that the constant holomorphic functions act like multiples of a delta
 function in Fock space: if $\omega_m$ has standard form $i
 dz^j\wedge d\overline z^j$, then the inner product on the local
 Fock space $\mathcal H_m$ also has standard form, and the inner
 product of a constant function $c$ with $\psi\in \mathcal H_m$
 gives
 \begin{equation}
     \langle c,\psi {\rangle_{}}_{\mathcal H_m} = c \int_{T_mM} \overline{\psi(z)}
           e^{-|z|^2/\hbar}d^nzd^n\overline z \\
    = \pi \hbar c \overline {\psi(0)}.
 \end{equation}

\section{The Full Quantization}\label{Sec:FullQuant}
\subsection{Pairing}

 Having constructed the quantum algebra $\mathcal A=(C^\infty(M),\ast)$ through
 deformation quantization, as well as the quantum Hilbert space
 $\mathcal H_P$ through geometric quantization, it is natural to ask
 whether one can define a representation of $\mathcal A$ on
 $\mathcal H_P$. If this is possible, the question arises whether
 one obtains the same operators for functions that are also quantizable in geometric
 quantization.
 As $\mathcal A$ and $\mathcal H_P$ depend on several parameters, the
 answer to the questions above might depend on them as well.

 Recall that in the two quantization schemes the
 following structures occur:
 \begin{center}
  \begin{tabular}{|l|l|}
    \hline
     geometric quantization &   deformation quantization \\ \hline
    metaplectic structure $\mathcal H\rightarrow M$ & metaplectic structure $\mathcal H\rightarrow
    M$ \\
    polarization $P$ &  symplectic connection $\nabla$  \\
    prequantum bundle $B\rightarrow M$; curvature $-\frac i\hbar \omega$
    & connection $D_\mathcal H(\nabla)$ on  $\mathcal H$; curvature $\frac i\hbar \omega$   \\
     line bundle $\Delta^P$ &\\
    \hline
  \end{tabular}
 \end{center}
 In the pure deformation case we did not know how to define states,
 because the connection on $\mathcal H$ was not flat. Now there is
 one obvious solution, to consider the product with the prequantum bundle: $\mathcal H\otimes
 B$. This is a Hermitian bundle with
  flat, metric connection, implying that
  \begin{equation}\label{korrHR1}
      \Gamma_{D}(\mathcal H\otimes B) := \{\psi \in
      \Gamma(\mathcal H\otimes B)\ |\ (D_\mathcal H\otimes \mathbf 1 + \mathbf 1\otimes
      \nabla^B)\psi=0\}
  \end{equation}
 defines a Hilbert space, as the scalar product of two flat sections does not
 depend on the base point:
  $$ d\langle \psi,\phi \rangle = 0 \qquad \forall \psi,\phi \in
  \Gamma_{D}(\mathcal H\otimes B).$$
 The observables remain unchanged, and act as $ \hat f
 \otimes \mathbf 1$ on $\mathcal H\otimes B$, where $\mathbf 1$ is the identity on $B$.
 Here it is important that the induced connection
 on End($\mathcal H\otimes B)$ is $ D\otimes \mathbf 1 +\mathbf 1\otimes d$
 (because $B$ is a rank one bundle, and therefore its
 connection form is central in End$(B)$),
 so that $\hat f\otimes \mathbf 1$ is parallel, and leaves the space (\ref{korrHR1}) invariant.

 However, the Hilbert space (\ref{korrHR1}) is not the one
 from geometric quantization, given as a subset of $\Gamma_P(\Delta^P\otimes
 B)$. Suppose there was a sesquilinear pairing
  $$\langle \cdot,\cdot\rangle : \Gamma_P(\Delta^P\otimes B) \times
  \Gamma_D(\mathcal H\otimes B) \rightarrow \mathbb C.$$
 Then we could define the quantum operator $\sigma(f)$ corresponding
 to $f\in C^\infty(M)$ through
  $ \langle \sigma(f) \psi , \phi \rangle = \langle \psi , (\hat
    f^\dagger\otimes \mathbf 1 )\phi \rangle .$
 Indeed there is a canonical pairing, its construction is most
 conveniently described separately for real and totally complex
 polarizations.

\paragraph{Real polarizations}
 We need the following definition, which has no counterpart in the
 complex situation:
\begin{defi}\label{compa_defi}
 A real polarization $P$ and a symplectic connection $\nabla$ are called
 compatible if
  \begin{equation}\label{compat_cond}
    \Gamma_{P}((\Delta ^P)^{-1} \otimes B)\times
    \Gamma_D(\mathcal H\otimes B) \overset{\langle\cdot,\cdot\rangle_{\mathcal H\otimes B}}
    \longrightarrow C^\infty_{P}(M).
  \end{equation}
 In general one expects the full space of functions
 $C^\infty(M)$ on the r.h.s.
\end{defi}
 We will discuss this condition at the end of the
 section, and assume it to be satisfied from now on. Then we can
 define a pairing
 \begin{align}\label{PairingDiffHilb}
   \langle\cdot,\cdot\rangle :\Gamma_P(\Delta^P\otimes B)&\times \Gamma_D(\mathcal H\otimes
   B) \rightarrow \Gamma_P\big((\Delta^P)^{-1} \otimes
   K^P \otimes B\big) \times
   \Gamma_D(\mathcal H\otimes B)    \rightarrow \mathbb C, \\
    \big(\sqrt{d^nq}\otimes \mu,\ &\psi\otimes \nu\big) \mapsto \big(\delta
    \otimes d^nq\otimes \mu,\ \psi\otimes \nu\big) \mapsto \int_{M/D}
    \langle \mu,\nu\rangle_B \langle \delta,\psi \rangle _{\mathcal
    H}d^n q \nonumber
 \end{align}
 In the first step $\delta \otimes \sqrt{d^n q}\in ( \Delta^P)^{-1}\otimes \Delta^P$ was
 inserted. We obtain
 \begin{equation}\label{PairingReal}
   \big\langle \sqrt{d^nq}\otimes \mu,\ \psi\otimes \nu \big\rangle=
   \int_{M/D} \mu\overline\nu\ \overline\psi(0) d^nq.
 \end{equation}
 The condition (\ref{compat_cond}) is necessary to ensure that $ \langle \mu,\nu\rangle_B
 \overline\psi(0)$ is polarized (does not depend on $p$), and thus can be integrated over $M/D$.

\paragraph{Complex polarizations}
 Here the situation appears to be simpler. We can identify
 $\sqrt{d^n z}$ with the locally defined constant section $c$ of $\Delta^P\subset\mathcal H$, whose value at
 any point is the constant wave function 1. Let $\pi$ be the orthogonal
 projection from $\mathcal H$ to $\Delta^P$, then we define

 \begin{align}
 \langle\cdot,\cdot\rangle :\Gamma_P(\Delta^P\otimes B)&\times \Gamma_D(\mathcal H\otimes
   B) \rightarrow \Gamma_P(\Delta^P\otimes B)
   \times \Gamma(\Delta^P\otimes B)\rightarrow \mathbb C,
   \\
  \big(\sqrt{d^n z}\otimes \mu,\ \psi\otimes \nu\big) \mapsto&
  \big(\sqrt {d^n z}  \otimes \mu ,\ \pi\psi\otimes
  \nu\big) \mapsto \int_M \langle \mu,\nu\rangle_B \langle
  c,\psi\rangle_\mathcal H (\det g)^{1/4} d^n z\wedge d^n\overline
  z\nonumber,
 \end{align}
 using the pairing $(\sqrt{d^nz},\sqrt{d^nz})\mapsto (\det
 g)^{1/4}d^nz \wedge d^n \overline z$ from geometric quantization. Up to normalization this gives
 \begin{equation}
  \big\langle \sqrt{d^n z}\otimes \mu,\ \psi\otimes \nu \big\rangle
  = \int_M \mu \overline \nu\ \overline  \psi(0) e^{-\frac K\hbar} (\det g)^{1/4}
   d^n z\wedge d^n\overline z.
 \end{equation}

\bigskip

 In the following we will also assume the pairing between our
 Hilbert spaces to be non-degenerate, in the sense that the map
  $  \mathcal H_P \rightarrow \Gamma_D(\mathcal H\otimes
  B)',\ \psi\mapsto \langle \psi,\cdot\rangle$
 is an isomorphism. This might impose a severe restriction on the admissible pairs of
 connections and polarizations, but it is difficult to find a more
 explicit formulation of this condition.

\begin{defi}
 A representation $\sigma$ of $(C^\infty(M),\ast)$ on $\mathcal H_P$
 is defined through
  \begin{equation}\label{physOp}
    \langle \sigma(f) \psi,\phi\rangle = \langle \psi
    ,(\hat f^\dagger \otimes 1) \phi\rangle \qquad \forall \psi \in \mathcal
    H_P,\ \phi \in \Gamma_D(\mathcal H\otimes B).
  \end{equation}
\end{defi}

 To be precise, some regularity conditions on $f$ will have to be
 imposed for the formula to make sense. In the flat case it would be
 enough to claim $f\in S'(\mathbb R^{2n})$ if one restricts the
 $\phi$s in (\ref{physOp}) to sections of the Schwartz bundle $\mathcal S\subset
 \mathcal H$, with fibre $\mathcal S(\mathbb R^n)\subset
 L^2(\mathbb R^n)$.

  We will see that on cotangent bundles our definition reproduces the
 geometric quantization operators. In this particular example one has
 a distinguished polarization, and it is possible to find a symplectic
 connection satisfying the compatibility condition.

 Formally the non-degeneracy condition on the pairing implies
 equivalence of the two algebra representations, so that
 nothing seems to be gained by the transition to $\mathcal H_P$. One
 has to remember however that the representation on
 $\Gamma_D(\mathcal H\otimes B)$ is in terms of infinite series in
 $\hbar$ and the local Weyl operators $y^\mu$, whereas on $\mathcal
 H_P$ one might hope to get an extension of geometric quantization
 by well-defined operators.

\subsection{Compatibility} \label{Sec_Comptbl} We want to discuss the
 implications of compatibility of a real polarization with a symplectic connection, in the sense of
 definition \ref{compa_defi}.
 Choose sections $\psi\in \Gamma_P\big((\Delta^P)^{-1}\otimes B\big)$ and $\phi \in
 \Gamma_D(\mathcal H\otimes B)$. From now on we denote
 $D_\mathcal H$ simply by $D$ as well, and write $D\otimes \nabla^B:= D\otimes \mathbf 1 +
 \mathbf 1\otimes \nabla^B$. Condition (\ref{compat_cond}) takes the form ($\forall X\in \Gamma(P))$
  $$ 0= X\big(\big\langle \psi,\phi\big\rangle_{\mathcal H\otimes B}
  \big) = \big\langle {(D\otimes \nabla^B)_{}}_X\psi,\phi\big\rangle _{\mathcal H\otimes B}+ \big\langle \psi,
  {(D\otimes \nabla^B)_{}}_X\phi\big\rangle_{\mathcal H\otimes B}.$$
 The last term vanishes, implying ${(D\otimes \nabla^B)_{}}_X\psi=0$. In other words
  \begin{equation}
      \Gamma_{P;\nabla}\big((\Delta^P)^{-1}\otimes B)\overset ! \subset \Gamma_{P;D\otimes
     \nabla^B}(\mathcal H\otimes B),
  \end{equation}
 where $\nabla$ is the partial connection.
 If $\psi=\delta \otimes \mu$ with $\delta\in\Gamma_{P;\nabla}\big((\Delta^P)^{-1}\big)$ and $\mu\in \Gamma_P(B)$
 this implies $D_X\delta=0$ for $X\in \Gamma(P)$, and therefore
 \begin{equation}\label{compat_cond2}
   \Gamma_{P;\nabla}\big((\Delta^P)^{-1}\big) \subset \Gamma_{P;D}(\mathcal
   H).
 \end{equation}
 Note that the 'integrability condition'
  $$ \big(D_XD_Y - D_YD_X -D_{[X,Y]}\big) \psi = \Omega(X,Y)\psi=0 \qquad \forall X,Y\in \Gamma( P)$$
 for $\psi\in \Gamma_{P;D}(\mathcal H)$ is satisfied due to the
 Lagrangian property of $P$, otherwise $\Gamma_{P;D}(\mathcal H)$ would contain
 only the zero section.
 Of course, the same applies to $\Gamma_P(B)$. If $\delta$ is in $\Gamma_{P;\nabla}\big((\Delta^P)^{-1}\big)$, then $
 X(\delta)=0$ for $X\in \Gamma(P)$, and (\ref{compat_cond2}) demands $A( X)\delta=0$,
 where $A$ is the connection form (\ref{FedconnectionForm}) of
 $D=d+A$. In Darboux coordinates adapted to $P$ our condition becomes:
  \begin{equation}\label{compat_cond3}
     \Big( \omega_{\overline ka}y^a  -\frac 1{2}
        \omega_{ab}\Gamma^b_{\overline kc}y^ay^c   -\frac 1{8} R_{abc\overline
        k}y^ay^by^c  + \dots\Big)\delta=0,
  \end{equation}
 where $\overline k$ denotes an index in $ P$, and the other ones are arbitrary. Due to the
 $\hbar$-grading all terms must vanish separately, in particular we
 have $\omega_{\overline ka}y^a \delta=0$, which is always satisfied
 due to $y\delta=0$ for $y\in  P^\flat$, cf. (\ref{KostantsConstr}). In the next higher order we get
 a condition on the symplectic connection: $$
    \Gamma_{\overline k ab}y^ay^b \delta=0 $$
 (recall that $\Gamma_{abc}$ is fully symmetric in its indices).
 This amounts to saying that $\Gamma_{\overline k\overline l a}=0$,
 and implies
  \begin{equation}
        \nabla \Gamma(P)\subset \Gamma(P\otimes T^\ast M).
  \end{equation}
 The third order gives
  $$ R_{abc\overline  k}y^ay^by^c \delta=0,$$
 implying that $R_{(abc)\overline k}$ vanishes if two or more of
 $a,b,c$ are in $P$. Here $(abc)$ denotes symmetrization in
 the indices. This condition however is automatically satisfied if
 the second order condition holds; the latter implies the vanishing
 of $R_{abcd}$ whenever three or more indices are in
 $P$, which can be seen from (\ref{curvatureFromConnection}). We
 must leave it as an open question whether the higher order conditions
 really impose further constraints on $\nabla$.

 A result of Bressler and Donin should be mentioned here \cite{bressler_PolDefQua}. They use a
 somewhat different formalism for deformation quantization, based on
 normal instead of Weyl ordering. Under the conditions
  $$ \nabla \Gamma(P)\subset \Gamma(P\otimes T^\ast M)$$
 and
  $$ R(X,Y)Z = 0 \qquad \forall X,Y,Z \in \Gamma(P)$$
 they show that the star product constructed from $\nabla$ satisfies the equation $f\ast
 g=fg$ for $f\in C^\infty_P(M)$ and arbitrary $g$ (which certainly does not hold for our star product; compare
 in the flat case).
 This strong result suggests
 that the two conditions should suffice to guarantee compatibility
 of $P$ and $\nabla$. They are both satisfied if the second order
 condition $\Gamma_{\overline {kl}a}=0$ holds, and also for the
 K\"ahler connection and holomorphic polarization, see below.

\section{Examples}
\subsection{The flat case with real polarization: $M=\mathbb
R^{2n}.$} We introduce linear
 coordinates $(q^i,p_j),\ i,j=1,\dots,n$, choose the
 constant symplectic form $\omega=dp_j\wedge dq^j$, the
 symplectic connection $\nabla=d$, and a real polarization given by
 $P=$ span$\{\partial_{p_j}\ |\ j=1,\dots,n\}$. The Fedosov
 connection on $\mathcal H$ is then
  $$ D= d+\frac i\hbar \big(dp_j \hat q^j - dq^j\hat p_j\big),$$
 and the operator $\hat f\in \Gamma_D(W)$ corresponding to $f\in
 C^\infty(M)$ is given by (\ref{vollerOperatorflach}), which is
 often written as $\hat f_{(q,p)}= f(q+\hat q,p+\hat p)$.

 Let us assume that the connection on the trivial prequantum bundle $B$
 has the form $\nabla^B = d-\frac i\hbar p_jdq^j$. Choose a constant section
 $\mu\neq 0$ of $B$. Then $\psi\otimes \mu$ is in $\Gamma_D(\mathcal
 H\otimes B)$ iff $\psi\in \Gamma(\mathcal H)$ satisfies the
 following equation:
  \begin{equation}\label{FlatC_flatWaveFct}
     \Big(d-\frac i\hbar p_jdq^j +\frac i\hbar \big(dp_j \hat q^j -
     dq^j\hat p_j\big)\Big)\psi=0.
  \end{equation}
 The value $\psi_{(q,p)}$ of $\psi$ in a point $(q,p)\in \mathbb
 R^{2n}$ is a wave function on $Q_{(q,p)} \subset T_{(q,p)} \mathbb
 R^{2n}$; we denote the variables in $Q_{(q,p)}\cong \mathbb R^n$ by $\tilde q=(\tilde q^1,\dots,\tilde
 q^n)$. The operators $\hat q^j$ and $\hat p_j$ in (\ref{FlatC_flatWaveFct}) act as
 multiplication with $\tilde q^j$ and differentiation $\frac \hbar i
 \partial_{\tilde q^j}$ respectively, and the
 solutions of equation (\ref{FlatC_flatWaveFct}) are given by
  \begin{equation}\label{FlatC_WF}
    \psi_{(q,p)}(\tilde q)= \chi(q+\tilde q)e^{-\frac i\hbar \tilde
    qp},
  \end{equation}
 for arbitrary functions $\chi\in L^2(\mathbb R^n)$. We want to determine
 the operator $\sigma(f)$ acting on the physical Hilbert space
 $L^2_P(\Delta^P\otimes B)$, defined by (\ref{physOp}). Assuming
 $\langle \mu,\mu\rangle \equiv 1$, insertion of (\ref{FlatC_WF})
 into the pairing (\ref{PairingReal}) results in
  $$ \langle a\sqrt{d^nq}\otimes \mu ,\psi\otimes \mu\rangle =\int_{\mathbb
  R^n}  a (q)\overline{\chi(q)}  d^n q. $$
 Observing that $q^j+\hat q^j$ acts on $\psi(\tilde q=0)$ as multiplication operator $q^j$,
 and $p_j+\hat p_j$ as differentiation $\frac \hbar
 i\partial_j$ applied to $\chi$, it is then obvious that $\sigma(f)$ is given by the well-known
 Schrödinger operator in Weyl ordering.
 In particular we have
  $$\sigma(q^j)\big(a\sqrt{d^nq} \otimes \mu\big)(q)= q^j\big(a\sqrt{d^nq} \otimes \mu\big)(q) \quad \text{and}\quad
  \sigma(p_j)\big(a\sqrt{d^nq} \otimes \mu\big)(q)=\frac \hbar i\big(\partial_j a \sqrt{d^n q}\otimes \mu\big)(q).$$

\subsection{The flat case with K\"ahler polarization: $M=\mathbb
C^n.$}\label{SSec_FlatKaehler}
 The quantization data are $\omega= idz^j\wedge dz^j$, corresponding to the K\"ahler potential
 $K(z,\overline z) =|z|^2$, $\nabla=d$, and
  $P=$ span$\{\frac\partial{\partial z^j}\ |\
 j=1,\dots,n\}$. We choose a gauge in which $\nabla^B= d-\frac
 1{\hbar}\overline z^adz^a$. The Fedosov connection on $\mathcal H$
 is
  $$ D= d + \frac 1\hbar\big(d\overline z^a \hat z^a - dz^a
  \hat{\overline z}^a\big).$$
 If $\mu\in \Gamma(B)$ is constant, then $\psi\otimes \mu$ is in
 $\Gamma_D(\mathcal H\otimes B)$ iff $\psi$ satisfies
 \begin{equation}
        \Big( d -\frac 1\hbar \overline z^adz^a + \frac 1\hbar\big(d\overline z^a \hat z^a - dz^a
  \hat{\overline z}^a\big)\Big) \psi=0.
 \end{equation}
 The solutions to this equation are given by
 \begin{equation}
  \psi_{(z,\overline z)}(\tilde z) = \chi(z+\tilde z) e^{-\frac
  1\hbar \overline z\tilde z},
 \end{equation}
 for arbitrary $\chi\in L^2_{hol}\big(\mathbb C^n,
 e^{-|z|^2/\hbar }d^n zd^n \overline z\big).$ Now one easily checks that $\overline z^a +
 \hat{\overline z}^a$ acts on $\psi$ as differentiation $\hbar
 \partial_{z^a}$ applied to $\chi$, and from the pairing between
 $\mathcal H_P= L^2_P(\Delta^P\otimes B)$ and $\Gamma_D(\mathcal H\otimes B)$:
  $$ \langle a\sqrt{d^n z} \otimes \mu,\psi\otimes \mu \rangle \sim
  \int_{\mathbb C^n} a(z) \overline
  {\psi(0)}|\mu|^2e^{-K/\hbar}d^nzd^n\overline z =
  \int_{\mathbb C^n} a(z) \overline {\chi(z)} e^{-|z|^2/\hbar} d^nzd^n
  \overline z $$
 it follows that one obtains the standard Fock space operators.

\subsection{Cotangent bundles} Let $(Q,g)$ be a (semi-)Riemannian
 manifold, and $M=T^\ast Q$. Suppose one has chosen local coordinates $q^j$
 on $Q$, then there is a set of canonical coordinates on $M$, given
 by $(q^j,p_j)$, where $p_i(dq^j) = {\delta_i}^j$. $M$ carries a
 canonical symplectic form $\omega=dp_j\wedge dq^j$, as well
 as a canonical polarization, given by the vertical subbundle of
 $TM$, i.e. $P_{(q,p)} = $
 span$\{\partial_{p_1},\dots,\partial_{p_n}\} \subset
 T_{(q,p)}T^\ast Q$. The prequantum bundle $B$ can be chosen to be
 trivial, as its curvature $\omega = d(p_jdq^j)$ is exact. It is not so obvious how to choose the
 symplectic connection. As we assumed $Q$ to be equipped with a
 metric, we have the Levi-Civita connection $\nabla^{LC}$ on $Q$. Bordemann et.al.
 have shown that there is a canonical lift of $\nabla^{LC}$ to $M$,
 which is symplectic and leads to a star product with nice
 properties \cite{BordemannNW-StarprodKot}. In particular the product of two functions polynomial
 in $p$ will be a finite polynomial in $\hbar$ and $p$, so that
 $\hbar$ need not be considered as a formal element in the end, but
 can be given its physical value.

\bigskip

 We will use the following notational convention. A latin index $i$
 runs from 1 to $n$ and stands for $q^i$, whereas a barred latin
 index $\overline i$ denotes $p_i$. Greek indices run from 1 to
 $2n$, and include all coordinates. Generators of the local Weyl
 algebra $W_m$ are denoted by $y^\mu$ (as in section \ref{sec_defQuant}), or by $\hat q^i$ and $\hat
 p_i$ (not to be confused with the elements of $\Gamma_D(\mathcal
 W)$ corresponding to the functions $q^i,p_i$. We never use the
 symbols $\hat q^i,\hat p_i$ in this latter sense, although it would be
 consistent with $f\mapsto \hat f\in \Gamma_D(\mathcal W)$).
 One forms are $dx^\mu$, or $dq^i$ and $dp_i$. The product
 $y^\mu y^\nu$ still denotes the symmetric tensor product, whereas $\hat
 q^i\hat p_i$ is composition of operators, or the Weyl product.

 The connection and curvature coefficients of the lifted connection on $M$ are
  \begin{align}\label{kotbdlIndZshg}
    \Gamma^{k}_{ij}&=-\Gamma^{\overline j}_{i\overline
       k}=-\Gamma^{\overline i}_{\overline kj}= \tilde\Gamma^k_{ij} ,\nonumber\\
    \Gamma^{\overline k}_{ij}=\frac {p^a}3& \Big(2\tilde
       \Gamma^a_{jl}\tilde\Gamma^l_{ki} - \partial_j\tilde
       \Gamma^a_{ki} + \text{cycl.}(ijk)\Big), \\
    R^l_{kij}=-R^{\overline k}_{\overline lij}&=\tilde
      R^l_{kij},\qquad R^{\overline l}_{k i \overline j} =\frac
      13\big(\tilde R^j_{lki}+\tilde R^j_{kli}\big), \nonumber\\
    R^{\overline i}_{jkl} = \frac {p^a}3 \Big( \nabla_i \tilde
      R^a_{jlk} - &3\tilde \Gamma^a_{im}\tilde R^m_{jlk} - \tilde
      \Gamma^a_{lm}\tilde R^m_{ijk} + \tilde\Gamma^a_{km}\tilde
      R^m_{ijl} + (i\leftrightarrow j)\Big),\nonumber
  \end{align}
  where $\tilde \Gamma^k_{ij}$ and $\tilde R^i_{jkl}$ are the
 Christoffel symbols and curvature tensor on $Q$, and $\Gamma^k_{ij},
 R^i_{jkl}$ the lifted objects on $T^\ast Q$. Components not
 listed vanish. Notice in particular that our first order condition
 $\Gamma_{\overline k\overline l m}=\Gamma_{\overline k\overline l \overline
 m}=0$ from (\ref{compat_cond3}) is satisfied.
 Now we want to show that the lifted
 connection $\nabla$ is compatible with the vertical polarization in
 every order in $\hbar$. We need the following
\begin{defi}
  For a homogeneous element
 \begin{equation}
  \eta= fy^{\mu_1}\dots y^{\mu_k} dx^{\nu_1}\wedge \dots\wedge
  dx^{\nu_l} \quad \in \Gamma\big(\text{Sym}_k(T^\ast M)\otimes
  \Lambda^l T^\ast M\big)
 \end{equation}
 $(f\in C^\infty(M))$ we define its $Q$-degree, $q(\eta)$, as the difference of the
 number of $\hat q^j$s minus the one of $\hat p_j$s, plus the number
 of $dq^j$s minus the one of $dp_j$s. The $Q$-degree defines a
 (vector space) filtration on $\Gamma(\mathcal W\otimes \Lambda T^\ast M)$; we
 write $q(\eta)>m$ if this holds for every homogeneous component of
 $\eta\in \Gamma(\mathcal W\otimes \Lambda T^\ast M)$.
\end{defi}
 Note that the $Q$-degree is not additive in general: $q(\eta\circ
 \chi) \neq q(\eta)+q(\chi)$, because of the relation $dq^i\wedge
 dq^i=0$. It is additive only when restricted to suitable sections:
 \begin{equation}
  q(\eta \circ \chi )=q(\chi\circ \eta) =q(\eta) +q(\chi),\quad \forall
  \eta \in \Gamma(\mathcal W),\chi\in\Gamma(\mathcal W\otimes
  \Lambda T^\ast M).
 \end{equation}
 Let us investigate the $Q$-degree of the connection form $A$ of $D$, and
 the operator in $\Gamma_D(\mathcal W)$ corresponding to a function.
 We use the notation of section \ref{sec_defQuant}.
\begin{lem}\label{QGradLemma}The following hold:
 \begin{itemize}
   \item $q(A)\geq 0$,
   \item $q\big(A-\frac i\hbar \omega_{\mu\nu}y^\nu dx^\mu\big)>0$,
     unless $\Gamma=0$,
   \item $q(r-r^{(k)})>k-1$ for $k\geq 3$, unless $r-r^{(k)}=0$.
   \item Let $f\in C^\infty(M)$ be a polynomial of degree $N$ in $p$, with
    arbitrary $q$-dependence. Then

     $q(\hat f-\hat f^{(k)}) > k-2N,$ unless $\hat f - \hat f^{(k)}=0$.
 \end{itemize}
\end{lem}
\begin{proof}
 The first two properties follow from the third, as the first terms
 in $A$, given by
  $$ \frac i\hbar \omega_{\mu\nu}y^\nu dx^\mu -\frac i{2\hbar} \Gamma_{\alpha\beta\gamma}y^\alpha y^\beta
  dx^\gamma$$
 have $Q$-degrees $=0$ and $>0$ respectively. This is because
 $\Gamma_{\alpha\beta\gamma}$ vanishes if two or more indices are
 barred, as can be read off from (\ref{kotbdlIndZshg}). The same
 holds true for $R_{\alpha\beta\gamma\mu}$, so that the third term
 $r^{(3)}$ in $A$
   $$  r^{(3)}= -\frac i{8\hbar} R_{\alpha\beta\gamma\mu}y^\alpha y^\beta y^\gamma dx^\mu  $$
 has $q(r^{(3)})>1$.
 Now we make use of the iteration formula (\ref{FedConnIteration}):
  $$r^{(n+1)}=\delta^{-1}\hat R +\delta^{-1}\big(\nabla r^{(n)}+\frac i\hbar (r^{(n)})^2\big) \mod
  \hbar^{(n+2)/2}$$
 for the higher order terms in $r$. Note that $\delta^{-1}$ leaves the $Q$-degree invariant.
 We have $\nabla = d+[dU(\Gamma),\cdot]$,
 where $d$ can only lower the $Q$-degree of the term proportional to
 $R_{ijkl}\hat q^i\hat q^j\hat q^kdq^l$ in $r^{(3)}$, which remains
 $>2$. The commutator $-\frac i\hbar
 \Gamma_{\alpha\beta\gamma}dx^\gamma [y^\alpha y^\beta,\cdot]$ increases
  the $Q$-degree by 1 (unless the result of its application
 vanishes), due again to
 $q(dU(\Gamma))\geq 1$. In the higher order iteration steps $\nabla$ can
 only decrease the $Q$-degree of terms containing either
 $R_{ijkl}$ or $\Gamma_{ijk}$ (only unbarred indices!), as these are linear in $p$ (cf. (\ref{kotbdlIndZshg})).
 These however must arise from terms $\Gamma_{ijk}\hat q^i\hat
 q^jdq^k$ or again the starting term $R_{ijkl}\hat q^i\hat q^j\hat
 q^kdq^l$, which both have a two units higher $Q$-degree as we would expect.
 Therefore we can infer that effectively every application of $\nabla$
 increases the $Q$-degree by 1, where we have still ignored the
 $(r^{(n)})^2$-term in the iteration formula. It is obvious from the (almost) additivity of the $Q$-degree
 that its inclusion does not change the picture.
 The same reasoning then works for $\hat f$, using the iteration
 formula (\ref{OperatorIterationGlg}), and the fact that $\nabla$
 can decrease the $q$-degree $N$ times.
\end{proof}

\begin{prop}
 In the situation of this paragraph we have $ \Gamma_{P;\nabla}\big((\Delta^P)^{-1}\big) \subset \Gamma_{P;D}(\mathcal
   H)$ (cf. \ref{compat_cond2}), thus the pairing (\ref{PairingDiffHilb}) is well defined.
\end{prop}
\begin{proof}
 As discussed above equation (\ref{compat_cond3}) we must show
 that $A_{(q,p)}(X)\delta = 0$ for every $X\in P_{(q,p)}$, where $\delta$ is
 the delta distribution on $Q_{(q,p)}\cong T_qQ$. This is the case
 iff every homogeneous term in $A_{\overline k}=A(\partial_{p^k})$ contains more operators $\hat q^j$ than
 $\hat p_j$s, i.e. if $q(A_{\overline k})>0$. But we have
 $q(A_{\overline k}) \geq q(A)+1 >0$ from the lemma above.
\end{proof}
 We conclude that the symplectic connection (\ref{kotbdlIndZshg}) is
 compatible with the vertical polarization of $T^\ast Q$, and
 (\ref{physOp}) defines an operator $\sigma(f)$ on the physical
 Hilbert space. Now we can compare our formalism to geometric
 quantization.

\begin{prop}\label{CotBdlOpsEqGeoOps}
 If $f$ is quantizable in geometric quantization, then
 $\sigma(f)$ coincides with the geometric quantization operator.
\end{prop}
\begin{proof}
 Let $f(q,p)=f(q)$ be a polarized function, and $g_j(q,p)= p_j$. From Lemma \ref{QGradLemma}
 we have
  \begin{equation}
    q(\hat f-f)>0,\qquad q(\hat g_j-\hat g_j^{(2)}) >0.
  \end{equation}
 It immediately follows that $(\hat f\psi)(0) =
 f\psi(0)$ for $\psi\in\Gamma(\mathcal H)$. Recalling the explicit solution
 (\ref{OperatorIteration})
  $$ \hat {g_{}}_j= {g_{}}_j+\partial_\mu{g_{}}_j y^\mu + \frac
  12\big(\partial_\mu\partial_\nu -
  \Gamma^\kappa_{\mu\nu}\partial_\kappa \big){g_{}}_j y^\mu y^\nu+
  O(\hbar^{3/2})$$
 to the iteration equation (\ref{OperatorIterationGlg}) for $\hat g_j$, we obtain
  \begin{equation}\label{affOpHKB}
    (\hat g_j\psi)(0) = \big(p_j + \hat p_j -\frac 12\Gamma^{\overline
    j}_{a\overline b}\hat p_b\hat q^a \big) \psi(0)
  \end{equation}
 This looks rather complicated yet, but will simplify in a moment.
 We want to apply the operators $\hat f\otimes 1,\ \hat g_j\otimes 1$ to elements of
 $\Gamma_D(\mathcal H\otimes B)$. Suppose that $\mu$ is a constant section
 of $B$. Then the condition for $\psi \otimes \mu$ to be in
 $\Gamma_D(\mathcal H\otimes B)$ is $ (d+A -\frac i\hbar
 \theta)\psi=0$, where $\theta=p_jdq^j$ is the connection form on
 $B$. But we know from Lemma \ref{QGradLemma} that $q\big(A-\frac i\hbar
 \omega_{ab}y^bdx^a\big)>0$, and $q(r)>1$ (unless $r$ vanishes).
 Therefore the condition evaluated in $\tilde q=0$ takes the simple
 form
  \begin{align}
   (d+A -\frac i\hbar \theta)(\partial_{p^j}) \psi(0) &=
    \partial_{p_j} \psi(0)= 0\\
   (d+A -\frac i\hbar \theta)(\partial_{q_j}) \psi(0) &=
   \Big(\partial_{q^j} - \frac i\hbar (p_j + \hat p_j) -\frac
   i{2\hbar} \Gamma_{ja\overline b }\hat p_b\hat q^a \Big)
   \psi(0)=0.\nonumber
  \end{align}
 The first equation tells us that $\psi(0)$ does not depend on $p$,
 ensuring that the integral in our pairing (\ref{PairingReal}) is well
 defined. Using $\Gamma_{ja\overline b} = -\Gamma^{\overline j}_{a\overline
 b}$ in Darboux coordinates, the second equation together with (\ref{affOpHKB}) gives
  \begin{equation}
    (\hat g_j\psi)(0)= \frac \hbar i \partial_{q^j} \psi(0).
  \end{equation}
 From the pairing $\langle\cdot,\cdot\rangle$ between
 $\Gamma_P(\Delta^P\otimes B)$ and $\Gamma_D(\mathcal H\otimes B)$
 \begin{equation}
   \langle a\sqrt{d^nq} \otimes \mu, \psi\otimes \mu \rangle =
   \int_Q  a(q) \overline{\psi_{q}(0)}d^n q
 \end{equation}
 (if $\langle \mu,\mu\rangle_B\equiv 1$) we then obtain the usual
 geometric quantization operators, i.e.
 \begin{align}
   \sigma(f)\big(a\sqrt{d^nq} \otimes \mu\big)(q) = f(q) a(q)\sqrt{d^nq} \otimes
   \mu,\\
   \sigma(p_j)\big(a\sqrt{d^nq} \otimes \mu\big)(q) =
   \frac \hbar i \partial_j a(q) \sqrt{d^nq}\otimes \mu.\nonumber
 \end{align}
 We still have to check that $\sigma\big(\alpha^j(q)p_j\big)$ gives the symmetrized
 product of $\alpha^j(\sigma(q))$ and $\sigma(p_j)$. But
 \begin{align*}
   \widehat{a^j(q)p_j}\psi(0) &= \Big[a^jp_j + \partial_k a^jp_j\hat
   q^k + a^j\hat p_j + \frac 12\big(\partial_l a^j -  a^k\Gamma^{\overline k}_{l\overline
   j}\big)  (\hat q^l\hat p_j +   \hat p_j\hat q^l)\Big]\psi(0) \\
    &= \Big[a^j\big(p_j +\hat p_j-\frac 12 \Gamma^{\overline j}_{l\overline
    k} \hat p_k\hat q^l\big) + \frac 12 \partial_l a^j\hat
    p_j\hat q^l\Big]\psi(0)\\
   &= \Big[a^j\hat g_j - \frac {i\hbar}2 \partial_j a^j\Big]\psi(0),
 \end{align*}
 which leads to the symmetrized product.
\end{proof}
 The next proposition is a pure deformation quantization result.
\begin{prop}
 Let $f$ and $g$ be polynomials of degree $N$ and $M$ in $p$
 respectively (with arbitrary dependence on $q$), and no $\hbar$-dependence. Then
 $f\ast g$ is a polynomial of degree $\leq N+M$ in $\hbar$.
\end{prop}
\begin{proof}
 Lemma \ref{QGradLemma} and the additivity of the $Q$-degree
 imply that $q\big((\hat f-\hat f^{(r)})\circ (\hat g - \hat g^{(s)})\big) > r+s-2(N+M)$.
 But we have $\pi_\otimes^0(\chi)=0$ if $q(\chi)>0$ for $\chi\in
 \Gamma(\mathcal W)$ ($\pi_\otimes^0$ projects onto the part
 containing no operators $\hat q,\hat p$, definition (\ref{GradProjektionen})).
 Therefore only terms $\hat f^{(r)}\circ \hat g^{(s)}$ with $r+s\leq 2(N+M)$ contribute to $f\ast g$
 (defined in (\ref{StarProdDefi})), but these have $\hbar$-degree
 $\leq \frac{2(N+M)}2=N+M$.
\end{proof}
 In fact a stronger result holds true \cite{BordemannNW-StarprodKot}; define
 $H=p_j \frac\partial{\partial p_j} + \hbar
 \frac\partial{\partial\hbar}$, then
  $$ H(f\ast g) = (Hf)\ast g + f\ast (Hg),$$
 a property called homogeneity in \cite{BordemannNW-StarprodKot}. Recall the
 expansion in $\hbar$ of the star product
  $$ f\ast g = fg + \sum_{k=1}^\infty \hbar^k c_k(f,g).$$
 We have shown that the series is indeed finite if $f$ and $g$ are
 polynomials in $p$. A monomial $h$ of degree $N$ in $p$ can be split
 as $h=h^jp_j$, where $h^j$ is a monomial of degree $N-1$. As our
 quantization map $\sigma$ is a homomorphism w.r.t. the star
 product, we obtain
  \begin{equation}
    \sigma(h)  = \sigma(h^j)\sigma(p_j) - \sum_{k=1}^N \hbar ^k
     \sigma\big(c_k(h^j,p_j)\big),
  \end{equation}
 enabling us to determine $\sigma(h)$ recursively, because all the
 functions on the r.h.s. are polynomials of degree at most $N-1$,
 and we know the operators corresponding to affine-linear functions.
 The calculations involved become very complex however, already for
 polynomials of degree 2! In this case the formula (for two
 affine-linear functions) reads
  \begin{equation}
    \sigma(fg) = \sigma(f)\sigma(g) - \frac
    {i\hbar}2\sigma\big(\{f,g\}\big) - \frac {\hbar^2}4
    \sigma\big((\nabla_\mu X_f)^\nu (\nabla_\nu X_g)^\mu \big).
  \end{equation}
 In \cite{Noelle_RelGEoDef} the operator corresponding to the kinetic energy
 term $g^{ab}(q)p_a p_b$ has been determined this way, the result being
  \begin{equation}\label{kinOpQuant}
    \sigma\big(g^{ab}p_ap_b\big)\big( \psi \otimes \sqrt{\mu}\big)
    = -\hbar^2\big[ \big(\Delta - \textstyle{\frac {\tilde R}4}\big) \psi\big] \otimes
  \sqrt\mu ,
  \end{equation}
 for $\psi\in \Gamma_P(B)=C^\infty(Q)$, $\mu=\sqrt{|\text{det } g|}d^nq$,
 where $\Delta$ is the Laplace-Beltrami operator, and $\tilde R$ the scalar curvature of
 $Q$. The section $\sqrt{|\text{det }g|}d^nq$ instead of $d^nq$ has been chosen to make the
 result independent of coordinates, so that Riemannian normal
 coordinates can be used, which simplify the calculation.

\paragraph{Comparison} A different quantization scheme for cotangent bundles exists,
 where the space of states is $L^2(Q,\sqrt{\det g}d^nq$) \cite{Landsman_Weyl}.
 It starts from the observation, that in the
 flat case the Schr\"odinger quantization map can be written as
  $$ \hat f\psi(x)= \frac{1}{(2\pi\hbar)^n } \int_{\mathbb R^{2n}} e^{\frac i\hbar  p(x-q)}
   f\big(\textstyle{\frac {x+q}2},p\big) \psi(q) d^n q d^np
      $$
 for $f\in \mathcal S'(\mathbb R^{2n})$, $\psi\in \mathcal S(\mathbb
 R^n)$. In order to generalize this formula to the curved case, we
 rewrite it in terms of the exponential function in the sense of
 differential geometry. Recall that in the flat case we have
 $T_x^\ast \mathbb R^n\cong \mathbb T_x\mathbb R^n \cong\mathbb R^n$ and
  $$ \exp_x : T_x\mathbb R^n \rightarrow \mathbb R^n,\ v\mapsto
  x+v,$$
 implying $ \frac {x-q}2 = \exp_{\frac {x+q}2}^{-1} (x). $
 Replacing $\mathbb R^n$ by a Riemannian manifold $Q$, we also have
 to replace the midpoint $\frac{x+q}2$ by the geodesic
 midpoint
  $$ m(x,q) := \pi\big(\nu^{-1}(x,q)\big),$$
 where
  $$        \nu: T Q \rightarrow Q\times Q,\ X\mapsto
  \big(\exp_{\pi(X)}(\textstyle{\frac 12} X), \exp_{\pi(X)}(-\textstyle{\frac 12}
  X) \big).$$
 Taking into account the measures $\sqrt{\det g(q)}d^n q$ on $Q$
 and $\frac{d^n p}{\sqrt{\det g(q)}}$ on $T_q^\ast Q$, one arrives at the formula
  \begin{equation}
    \hat f\psi(x) = \frac1{(2\pi\hbar)^n}\int_Q d^nq\int_{T_{m(x,q)}^\ast Q} d^np
    \textstyle{\sqrt{\frac{\text{det }g(q)}{\text{det }g(m(x,q))}}}
    e^{\frac{2i}\hbar p\big(\exp_{m(x,q)}^{-1}(x)\big)}
    f\big(m(x,q),p\big)\psi(q).
  \end{equation}
 This is well defined for functions $f$ that are polynomial in $p$, in
 general one has to include a cutoff function, as $\nu$ is not a
 global diffeomorphism $TQ \cong Q\times Q$. The details can be
 found in \cite{Landsman_Weyl}, where also the operators corresponding to functions
 $q^i,p_j$, and $g^{ab}(q)p_ap_b$ have been calculated in this formalism, with the
 results
  \begin{align}
        \hat q^i &= q^i, \nonumber\\
        \hat p_j &= -i\hbar \big(\partial_j + \textstyle{\frac 12} \Gamma^k_{kj}\big),    \\
        \widehat{g^{-1}(p,p)}& = -\hbar^2 (\Delta-\textstyle{\frac
        R3}).\nonumber
  \end{align}
 The first two operators again coincide with the ones from geometric
 quantization if $\sqrt{\sqrt{\det g}d^n q}$ is chosen as constant section of
 $\Delta^P$. The kinetic energy operator however is different from the
 one we obtained (\ref{kinOpQuant}).

 It appears to us as a drawback of the above quantization scheme
 that the metaplectic group, which is the symmetry group of quantum mechanics on a symplectic
 vector space, does not seem to play any role there. But there is no criterion to decide
 which quantization is the physical one.

\subsection{A remark on K\"ahler manifolds}
 On a K\"ahler manifold one has the holomorphic polarization, and also
 a canonical symplectic connection, the Levi-Civita
 connection of the K\"ahler metric $g$. Its
 nonvanishing components are \cite{jost:RiemGeom}
  $$ \Gamma_{\overline a bc} = -i\partial_{\overline
  a}\partial_b\partial_c K,\qquad \Gamma_{a\overline b\overline c} =
  i\partial_a\partial_{\overline b}\partial_{\overline c}K.$$
 We are interested in the operators corresponding to holomorphic
 functions, as well as the function $\frac \partial {\partial
 z^a}K$, in order to compare to the geometric quantization operators
  (\ref{GeomQuantKaehlerOp}). In the case of cotangent bundles we made use of the
 fact that the covariant derivative $D$ on $\mathcal H$ contained
 only finitely many terms contributing to $D\psi(0)$, which made it possible to
 determine the operator corresponding to
 $p_a$ in the proof of proposition \ref{CotBdlOpsEqGeoOps}.

 There seems to be no analogous property of $D$ when constructed from a
 K\"ahler connection, and the method from above cannot be applied
 here. We take a look at the first terms appearing in the
 corresponding expressions nevertheless. Let $\mu\in \Gamma(B)$ be
 constant (locally, in a trivialization with $\theta=-i\partial K$), then
 $\psi\otimes \mu$ is in $\Gamma_D(\mathcal H\otimes B)$ iff
 $(D-\frac i\hbar \theta)\psi=0$. We calculate
  \begin{align}\label{KaehlerFlatness}
           \big(D -\frac i\hbar \theta\big)_{\frac \partial{\partial z^a}}\psi(0)
    &= \Big[ \frac \partial{\partial z^a} - \frac 1\hbar \partial _a
      K -\frac 1\hbar \partial_a\partial_{\overline b}K \hat
      {\overline z}^b- \frac 1{4\hbar}
      \partial_a\partial_b\partial_{\overline c}K \hat {\overline
      z}^c \hat z^b+ \mathcal O(\hbar^{1/2}) \Big]\psi(0),\nonumber \\
   \big(D -\frac i\hbar \theta\big)_{\frac \partial{\partial \overline z^a}}\psi(0)
    &= \Big[ \frac \partial{\partial\overline z^a} +\frac1{4\hbar} \partial_{\overline
       a}\partial_b\partial_{\overline c} K \hat {\overline
       z}^c \hat z^b+ \mathcal O(\hbar^{1/2}) \Big]\psi(0).
  \end{align}
 In the real case the second equation was $(D-\frac i\hbar
 \theta)_{\frac\partial {\partial p}} \psi(0)= \frac
 \partial{\partial p} \psi(0)=0$, and told us that $\psi(0)$ was
 independent of $p$. One would expect to find $\psi(0)$ being
 holomorphic here, instead we have
  $$ \frac \partial{\partial \overline z^a} \psi(0 ) =- \frac i4
  \Gamma^{\overline c}_{\overline c\overline a}\psi(0) +\mathcal O(\hbar^{1/2}).$$
 Further, the two equations (\ref{KaehlerFlatness}) do not seem to
 imply any simplifications on the operators $\hat f^\dagger=\hat {\overline f}$ and
 $\widehat{\partial_a K}^\dagger=\widehat {\partial_{\overline
 a}K}$, for holomorphic $f$, which are needed to determine
 $\sigma(f)$ and $\sigma(\partial_a K)$.

 There is still an obvious necessary condition for the results $\sigma(f)=f$ and $\sigma(\partial_a
 K)=\hbar\frac\partial {\partial z^a}$, as well as symmetrized
 operators for products $f\partial_a K$. For the star product these
 would imply the relations $f\ast g=fg$ for holomorphic $f,g$, and
  \begin{equation}\label{KaehlerSeriesCond}
        f\ast \partial_a K = f\partial_aK +
 \frac{i\hbar}2\{f,\partial_a K\},
  \end{equation}
 which can be checked in pure deformation quantization. There is no
 obvious way however to calculate the star products exactly. A term by term
 calculation up to $\mathcal O(\hbar^4)$ suggests that they might indeed
 hold \cite{Noelle_RelGEoDef}.

 One should also keep in mind that it is not strictly necessary to
 work with the K\"ahler connection here, in principle any symplectic
 connection is possible. But the
 condition (\ref{KaehlerSeriesCond}) will certainly not be
 satisfied for arbitrary connections, it is indeed very restrictive
 and might even single out the K\"ahler connection. It is also
 questionable whether the pairing we constructed is always
 non-degenerate.

\bigskip

 After all there is not much we can say about K\"ahler manifolds,
 except in the trivial case $M=\mathbb C^n$. This might be
 an indication that the presented formalism needs some modification
 in general.

\section{Conclusion}
\parindent=1em

 We have shown that it is possible to define a representation of the
 deformation quantization algebra on the Hilbert space of geometric
 quantization, if some conditions are satisfied. For real
 polarizations one has to find a compatible connection (definition
 \ref{compa_defi}), and in general the pairing of the different
 Hilbert spaces should induce an isomorphism between them.

 The difficulties in geometric quantization to define operators of
 higher order in $p$ or $\overline z$ can be explained in this
 formalism by the fact that they depend on the chosen connection,
 and are therefore not uniquely determined by the geometric
 quantization data $(\omega,P,B,\mathcal H)$.

  An open problem is the determination of the quantum operators on a K\"ahler
 manifold. If this can be solved, the resulting theory should be
 compared to existing quantization schemes on K\"ahler manifolds, like Berezin, Berezin-Toeplitz
 (these are generalizations of normal and antinormal ordered
 quantization, and we do not expect our quantization to coincide
 with either of them) and of course geometric quantization. See
 \cite{Ali_QuantMeths} and references therein for an overview of different
 quantization schemes.

 When should we consider two quantizations $(\mathcal A_1,\mathcal H_1)$ and $(\mathcal A_2,\mathcal
 H_2)$ of $(M,\omega)$ as equivalent? Of course, they must give the
 same expectation values for the observables. Suppose there is an
 algebra isomorphism $\alpha: \mathcal A_1\rightarrow \mathcal A_2$
 and a unitary map $U:\mathcal H_1\rightarrow \mathcal H_2$ such that
 $\alpha(f)=UfU^{-1}$. Then the two quantizations are equivalent if we are willing to identify $\psi\in
 \mathcal H_1$ with $U\psi\in \mathcal H_2$ and $f\in \mathcal A_1$
 with $\alpha(f)\in \mathcal A_2$, because
  \begin{equation}\label{concl_equivalence}
        \langle \psi,f\psi \rangle_1 = \langle
  U\psi,\alpha(f)U\psi\rangle_2.
  \end{equation}
 This is an appropriate mathematical notion of equivalence,
 physically however the elements of the algebra are fixed
 observables (recall that as sets our algebras are contained in
 $C^\infty(M)$), and we cannot identify $f$ with the different
 observable $\alpha(f)$.

 If an algebra is compatible with two different polarizations, we
 still want to identify the two systems, so if the corresponding
 representations are $\sigma_i:\mathcal A\rightarrow $End($\mathcal
 H_i)$, $i=1,2$, and $U:\mathcal H_1\rightarrow \mathcal H_2$ is a
 unitary map satisfying
  \begin{equation}\label{concl_equivalence2}
        \sigma_2(f)=U^{-1}\sigma_1(f)U
  \end{equation}
 then we consider $(\mathcal A,\mathcal H_1$) and $(\mathcal A,\mathcal
 H_2$) as equivalent, and identify $\psi\in \mathcal H_1$ with
 $U\psi\in \mathcal H_2$. Note that $\sigma_i$ are really the
 particular representations constructed above, and determine $U$ uniquely up to a phase.

 Another way to see that not all quantizations satisfying (\ref{concl_equivalence})
 are equivalent, is to consider a fixed polarization with two
 different algebras. We know that $|\psi|^2$ is supposed to have a physical meaning as a
 probability density on $Q=M/D$, and unless $U\in U(1)$ we cannot
 identify $\psi$ with $U\psi$, because they define different
 densities. Therefore we must consider any two quantum systems
 constructed from different symplectic connections as physically inequivalent.

 What is then the moduli space of inequivalent quantizations of
 $(M,\omega)$? Assuming that we can always find a unitary
 intertwiner $U$ satisfying (\ref{concl_equivalence2}), the
 polarization is determined by the connection up to equivalence
 (however, there are symplectic manifolds not admitting any
 polarization at all, which are therefore not quantizable \cite{gotay_NonexistencePol}).
 There is then still a dependence on the prequantum bundle $B$, the
 metaplectic structure $\mathcal H$, and the symplectic connection
 $\nabla$. In the example of cotangent bundles we have seen that
 different symplectic connections correspond to physically
 inequivalent classical systems. The same is true for the prequantum
 bundle $B$, but the interpretation of the metaplectic structure is
 not obvious. Neither do we understand the meaning of the
 symplectic connection in general. It is also very questionable whether
 one can always find a unitary intertwiner $U$ as in
 (\ref{concl_equivalence2}) for different polarizations compatible
 with a given connection, because the set of compatible
 polarizations contains at least all totally complex polarizations.
 A possible way out might be to impose a compatibility
 condition on these as well, like the one obtained by Bressler and
 Donin \cite{bressler_PolDefQua}, see our discussion at the end of
 section \ref{Sec_Comptbl}, or to require holomorphic functions to be quantized to
 multiplication operators. Further one needs the pairing between the
 different Hilbert spaces to be non-degenerate, which might also
 restrict the admissible polarizations. The construction of $U$ is discussed in
 \cite{woodhouse:quantisierung}.

\section*{Acknowledgement}
 I would like to thank my former advisor G. Grensing for many discussions
 on the subject.

\bibliographystyle{Lit}
\bibliography{literatur}

\end{document}